\documentclass[10pt, a4paper, onecolumn]{googledeepmind}
\usepackage{graphicx}
\usepackage{amsmath}
\usepackage{booktabs}
\usepackage{longtable}
\usepackage{multirow}

\usepackage{todonotes}

\usepackage{tabularx}
\usepackage{hyperref}
\usepackage{tikz}
\hypersetup{
	colorlinks   = true, 
    urlcolor     = black, 
	linkcolor    = gray, 
	citecolor   = gray 
}

\usepackage{xparse}
\usepackage{comment}
\usepackage{algorithm}
\usepackage{algpseudocode}
\usepackage[capitalise]{cleveref}
\usepackage{subcaption}
\usepackage[dvipsnames]{xcolor}
\usepackage{overpic}
\usepackage{fontawesome}

\newcommand{\attitudinal}{\textcolor{Magenta}{\footnotesize\faCircle}  }
\newcommand{\trajectory}{\textcolor{Dandelion}{\scalebox{1.2}{$\blacktriangleright$}}  }
\newcommand{\interaction}{\textcolor{Cerulean}{$\blacksquare$}  }

\newcommand{\usesubsubsection}{0}  
\newcommand{\mysub}[1]{%
  \ifnum\usesubsubsection=1
    \subsubsection{#1}%
  \else
    \noindent\textbf{#1:}%
  \fi
}

\begin{document}

\title{Linking Behaviour and Perception to Evaluate Meaningful Human Control over Partially Automated Driving}

\author[1,2,*,+]{Ashwin George}
\author[1,3,*]{Lucas Elbert Suryana}
\author[1,2]{Lorenzo Flipse}
\author[1,3]{Bart van Arem}
\author[1,2]{David A. Abbink}
\author[1,3]{Simeon Craig Calvert}
\author[1,4]{Luciano Cavalcante Siebert}
\author[1,2]{Arkady Zgonnikov}

\affil[1]{Centre for Meaningful Human Control}
\affil[2]{Department of Cognitive Robotics}
\affil[3]{Department of Transport \& Planning}
\affil[4]{Department of Intelligent Systems; Delft University of Technology}
\affil[*]{Authors contributed equally and listed in alphabetical order}
\affil[+]{Corresponding author: \url{A.George@tudelft.nl}, Mekelweg 2, 2628 CD, Delft, Zuid Holland, The Netherlands.}

\begin{abstract}
Partial driving automation creates a tension: drivers remain legally responsible while being less active in control.
Meaningful human control (MHC), a normative framework that can potentially address this tension, proposes that automated systems are designed to track relevant human reasons and that humans should at all times remain in control and be responsible.
However, empirical methods for evaluating whether systems are under MHC remain underdeveloped. 
In this driving simulator study, we investigated the extent to which 24 drivers experienced MHC when interacting with partially automated driving systems under two modes --- haptic shared control and traded control.
During overtaking manoeuvres on a two-lane, two-way road with fully automated longitudinal control and partially automated lateral control, drivers' actions were necessary to prevent crashes due to silent automation failures.
Starting from hypotheses derived from the properties of systems under MHC, we used a mixed-methods approach that links behavioural metrics, subjective post-trial ratings, and qualitative feedback to assess drivers' perception of responsibility and control. A confirmatory analysis indicated a negative correlation between the perception of the automated vehicle understanding the driver and conflict in steering torques.
Qualitative feedback revealed that mismatches in intentions between the driver and automation, lack of safety, and resistance to driver inputs reduced perceived MHC, while subtle haptic guidance aligned with driver intent had a positive effect. Thus, future designs should prioritise effortless driver interventions, transparent communication of automation intent through haptic, visual or auditory cues, and clear authority allocation to strengthen meaningful human control in partially automated driving.
\end{abstract}

\maketitle

\section{Introduction}
\label{sec:Introduction}

Responsibility goes hand-in-hand with the level of control~\cite{flemischDynamicBalanceHumans2012}. For example, the driver of a car has a much higher degree of responsibility and control over the vehicle
as compared to a passenger in a train. This balance of responsibility and control can be disrupted when automation systems take up part of the control while society still holds the drivers of such vehicles responsible for accidents~\cite{beckersDriversPartiallyAutomated2022}. Therefore, understanding driver's perception of responsibility and control in partially automated vehicles is essential when designing automated driving systems.

Vehicle automation is widely seen as a promising way to improve road safety and traffic efficiency~\cite{fagnant2015}. Early deployments of fully automated systems have demonstrated benefits, such as reductions in specific crash types \cite{kusano2025waymo}. In practice, however, most commercially available driving automation systems today operate at SAE Level 2 \cite{kim2026comparative}. According to SAE J3016 \cite{sae2021taxonomy}, Level 2 systems automate lateral and longitudinal vehicle control, while the human driver must continuously supervise the driving environment and remain responsible for the dynamic driving task. Consequently, manoeuvres such as overtaking remain the driver's responsibility, even when automation assists with vehicle control. These automation levels create a paradox: drivers are expected to remain constantly attentive even though their control role is significantly reduced \cite{endsley2017autonomous}.

One manifestation of this paradox is driver complacency, where over-reliance on automation reduces vigilance and increases reaction times. This decline in performance is associated with the decrement in vigilance and engagement caused by the cognitive underload when drivers shift from active control to passive monitoring~\cite{chu2023automation,endsley2017autonomous, merat2014human}. Such disengagement can have severe consequences in emergency scenarios where drivers have to take back control on short notice. 
The risk posed by the paradox of driver complacency is exacerbated by some companies marketing their products as being capable of ``autonomous driving'' while still relying on drivers to intervene when needed. When automated systems rely on human intervention for ensuring safe operation without properly engaging drivers, this raises questions regarding the legal and moral responsibility ascribed to drivers of automated vehicles.

Even though drivers remain legally accountable,
partial automation can erode drivers’ sense of responsibility for vehicle behaviour.
This reduced sense of responsibility is often explained through the \textit{sense of agency} (SoA), the subjective experience of initiating and controlling an action. Prior work shows that SoA tends to decline as automation increases, with higher autonomy reducing awareness and engagement \cite{moore2016sense, cornelio2022sense, berberian2012automation}. When individuals feel they did not cause an action, they are also less likely to accept moral responsibility for its outcomes \cite{moretto2011experience}.

Taken together, complacency and declining SoA create a layered challenge: systems expect drivers to remain attentive and accountable, yet their design undermines the very capacities required to do so. This mismatch produces misattribution of responsibility \cite{matthias2004responsibility, santoni2021four}, where it is unclear who should be responsible (and held accountable) for the actions of an automated system. Recent evaluations of commercial driving systems illustrate these gaps: drivers often lack understanding of how well the system can react (or how well they themselves can react) while manufacturers frequently distance themselves from responsibility through disclaimers \cite{calvert2020gaps}. 

\begin{figure}[t]
    \centering
    \includegraphics[width=0.8\linewidth]{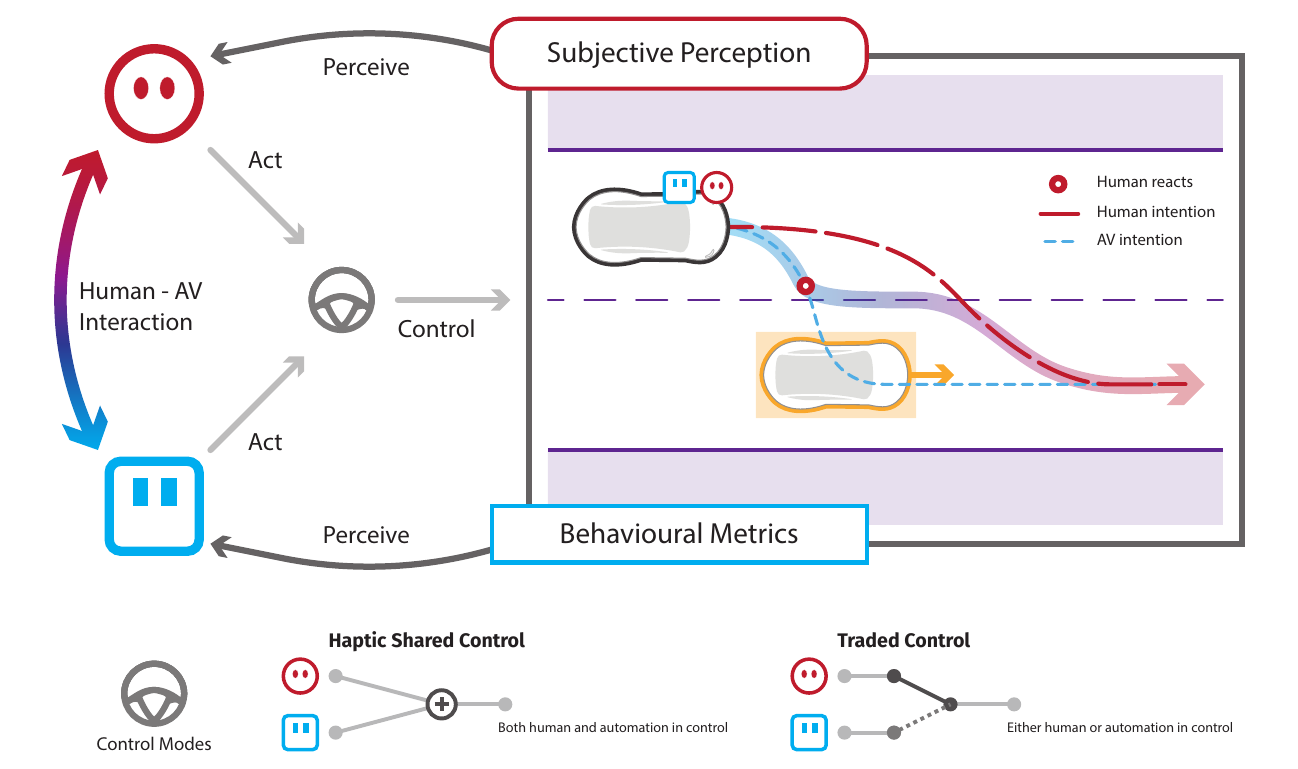}
    \caption{
    Understanding the relationship between the subjective perception of human drivers and behavioural metrics observable to machines is vital for designing automated systems under meaningful human control.
    We propose a mixed-methods approach to evaluate drivers perception of meaningful human control (MHC) when interacting with an automated driving system during overtaking manoeuvres on a two-lane, two-way road with fully automated longitudinal control and partially automated lateral control.
    This study considers two control modes --- haptic shared control (HSC) where the driver continuously corrects the automation and traded control (TC) where the human intervenes when necessary. During the task, silent automation failures will lead the partially automated ego vehicle (shown in black) to follow trajectories (shown in cyan) that crash into neighbouring vehicles (shown in yellow). The drivers' actions (shown by the red dot) are necessary to prevent crashes. Thus the collective behaviour of the ego vehicle is a combination of the automation reference (shown in blue) and the ideal behaviour of the human driver (shown in red). 
    }
    \label{fig:visual_abstract}
\end{figure}

One approach to mitigate such responsibility gaps would be to design automation systems based on the principles of meaningful human control (MHC). MHC has been proposed as a normative stance that humans should maintain some form of control and responsibility over the behaviour of artificial systems, even if these systems perform tasks autonomously~\cite{santoni2018meaningful}. 
Automated systems (including driving automation) must meet two conditions to be under MHC --- 1) the \emph{tracking condition} which asserts that the behaviour of automated systems must track the reasons of relevant humans and 2) the \emph{tracing condition} which asserts that it should be possible to trace the responsibility for the behaviours of the automated system to at least one human~\cite{santoni2018meaningful}. 
Based on these two fundamental conditions, subsequent works have proposed a range of practice-oriented frameworks to operationalize tracking and tracing, such as the four actionable properties of Cavalcante Siebert et al.~\cite{cavalcantesiebertMeaningfulHumanControl2022} or MHC operationalisation framework of Calvert~\cite{calvert2025principles}. 
The main idea of MHC is that, unlike the use of simple tools where the user of a tool can be held solely responsible, the use of automated systems might introduce responsibility gaps which might need to be addressed by the proper design of transition of control and responsibility across different levels in the chain of design and operation. For example, in the case of automated driving, responsibility and control could be dispersed among the government, regulatory bodies, automation manufacturers, and the driver of the automated vehicle~\cite{calvertDesigningAutomatedVehicle2023}. Depending on how these systems are designed, they could respond to distal reasons of humans higher up the chain of design and operation or be more responsive to the proximal reasons of the driver~\cite{mecacci2020meaningful}. Ideally, within their moral operational design domain (moral ODD) these automated systems should be designed to track the reasons set by humans, and departures from their moral ODD should be dealt with proper transfer of control and responsibility to human actors~\cite{cavalcantesiebertMeaningfulHumanControl2022,vanderwaaAllocationMoralDecisionMaking2020}. For this study, we focus on only one human, the driver of a partially automated driving system and their perception of MHC.

While MHC has often been discussed as a guiding principle for system design \cite{santoni2018meaningful, mecacci2020meaningful, cavalcantesiebertMeaningfulHumanControl2022, calvert2025principles}, a necessary first step is to evaluate whether an existing system satisfies its conditions in practice. Without such evaluation, it remains unclear whether MHC can be achieved in practice. Even though previous studies have proposed several ways to evaluate MHC of specific systems~\cite{calvert2020gaps, suryana2025meaningful}, they assess it indirectly
through post-crash analyses \cite{calvert2020gaps} or user interviews \cite{suryana2025meaningful}. Evaluations based on post-crash analyses do not capture the nuances of real-time interaction between the driver and the automated system. 
Although partial alignment with MHC principles can be inferred in case of 
prior evaluations based on user interviews,  the questions  were not specifically designed to measure MHC.
More generally, the current literature lacks methods that directly evaluate whether the interaction between human drivers and automated driving systems supports MHC.

In this work, we address this gap by combining subjective and behavioural measures to evaluate meaningful human control in driver interactions
with automated driving systems. Our evaluation methodology (\cref{fig:visual_abstract}) integrates behavioural metrics from driving interaction data and subjective questionnaires based on the established MHC properties~\cite{cavalcantesiebertMeaningfulHumanControl2022} to explicitly assess drivers' perception of MHC. This approach is inspired by Verhagen et al. \cite{verhagen2024meaningful}, who combined subjective reports and interaction metrics to evaluate MHC in human-robot interaction for firefighting systems. 
To demonstrate our evaluation methodology, 
we use two control modes based on haptic shared control (HSC) and traded control (TC), as representative control strategies.  
In HSC, both the driver and the automation act simultaneously through force feedback on the steering wheel, and the driver continuously corrects the behaviour of the automation~\cite{abbink2018topology, wang2017effect, li2018shared}. In TC, full control authority shifts between the driver and the automation based on whether the human is intervening by applying torques on the steering wheel~\cite{deWinter2023shared}. Our implementations of HSC and TC were chosen to contrast the continuous correction of HSC with the intermittent interventions of TC and does not represent state-of-the-art implementations.
With these two interaction strategies in mind, we address the following research questions (RQ):
\begin{enumerate}
\item [\textbf{RQ1}] \textit{To what extent do stereotypical strategies of human-automation interaction enable drivers to experience meaningful human control?}
\item [\textbf{RQ2}]
\textit{How are objective metrics associated with driver's subjective experience of meaningful human control?}
\end{enumerate}

We investigate these questions in a controlled driving simulator study where participants interact with an automated vehicle under both HSC and TC conditions  in safety-critical situations during overtaking manoeuvres on a two-lane, two-way road with fully automated longitudinal control and partially automated lateral control. The main contributions of this paper are:
\begin{itemize}[topsep=0pt]
    \item \textbf{Methodological contribution:} A methodology that integrates behavioural metrics, subjective questionnaires, and qualitative insights to evaluate how meaningful the control of human drivers is when interacting with driving automation 
    \item \textbf{Empirical comparison of two control modes:} An experiment-based comparison of two established human-automation interaction strategies in terms of the driver's perception of control and responsibility and the relation of those perceptions to observable behaviour.
    \item \textbf{Broader MHC contribution:} Implications for the design and evaluation of driving automation systems that aim to support the four properties of MHC.
\end{itemize}

\begin{figure}[tb]
\centering
    \includegraphics[width=0.8\linewidth]{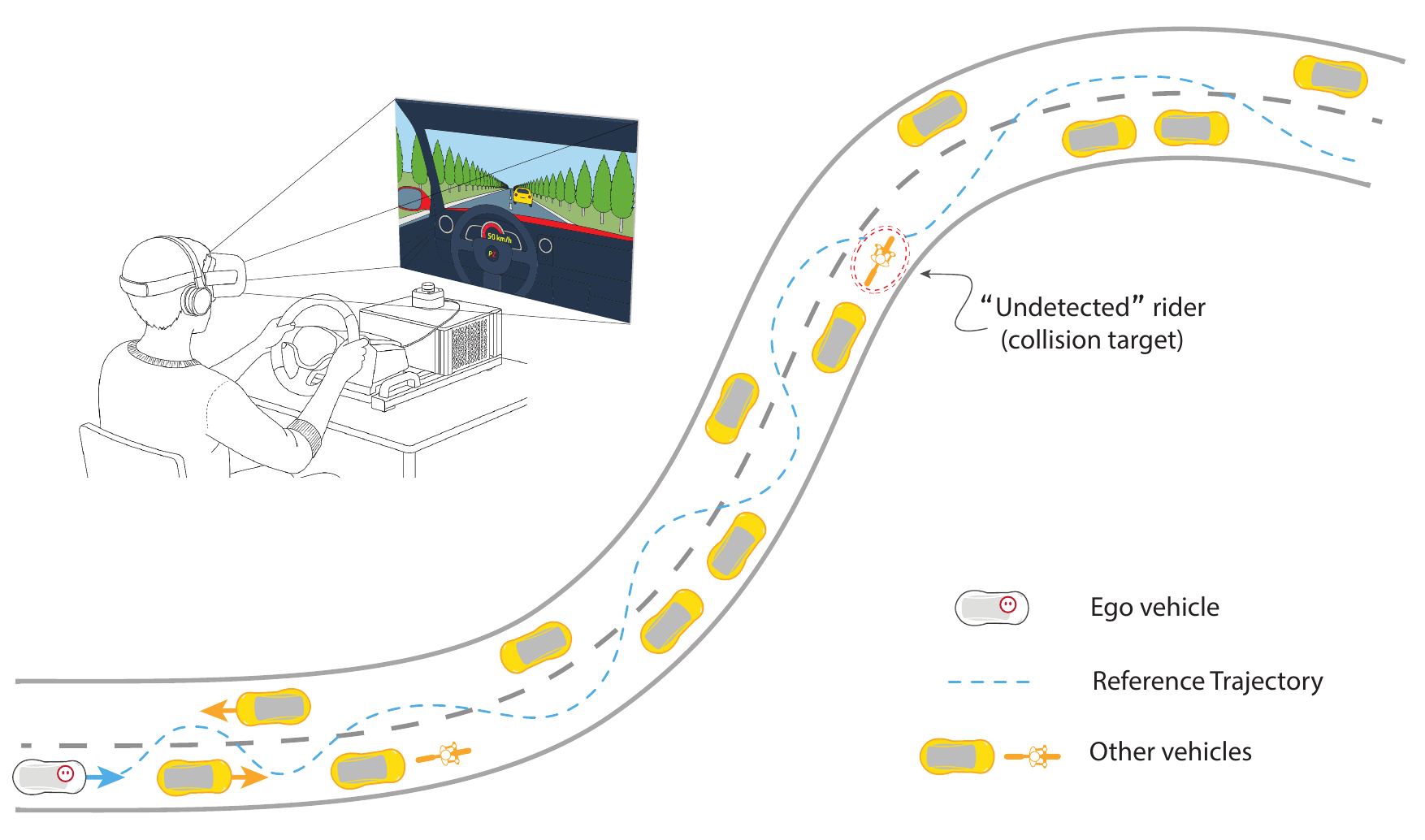}
    \includegraphics[width=0.9\linewidth]{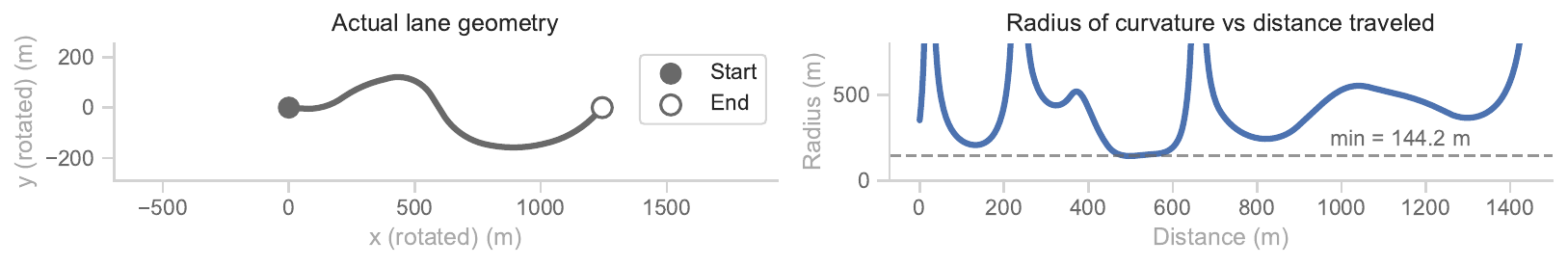}
\caption{\textbf{Driving simulator experiment:} Experimental setup: the fixed-base driving simulator with a virtual reality headset and haptic steering wheel, and the repeated overtaking scenario simulating a silent automation failure on a two-lane, two-way road lined with trees.
The actual lane geometry and the radii of curvature along the path are shown in the bottom panels.
The ego vehicle (white car) overtakes other vehicles and motorcyclists in the presence of oncoming traffic (yellow). 
To simulate silent automation failure (i.e., the system failing to detect a motorcyclist), the reference trajectory was used that crossed the path of a motorcyclist, leading to a collision in case the participant did not intervene. Only one motorcyclist served as the collision target, while the other travelled without conflict. The figure illustrates one representative configuration; the position of the vehicles and motorcycles, including the target motorcycle, were
counterbalanced across trials using a Latin square design across trials. 
}
\label{fig:experimental_design}
\end{figure}

\section{Methods}

We conducted a driving simulator experiment (\cref{sec:driving_experiment}) and used surveys to gauge  
drivers' subjective perception of interacting with driving automation through different control modes (\cref{sec:surveys}).
These subjective measurements were complemented with objective behavioural metrics derived from the telemetry data (\cref{sec:behavioural_metrics}), which were further analysed to study the influence of control modes, and relations between subjective perceptions and behavioural metrics (\cref{sec:analysis}). The data and analysis scripts are publicly available along with supplementary materials\footnote{Link to supplementary materials: \url{https://osf.io/rd53c/overview?view_only=13df8140aecf463abf2588fe018b434b}}.

Twenty-four participants (13 male, 11 female) 23 to 36 years old (with average and SD 29.2 ± 3.75 years) were recruited from the student and research community at Delft University of Technology between December 2023 and April 2024 via flyers distributed through personal contacts and snowball sampling, where enrolled participants referred others.
Eligible candidates must (i) have held a valid driving licence for at least one year, (ii) have normal or corrected-to-normal vision without spectacles (contact lenses were permitted), and (iii) have no history of epilepsy or other conditions that could be aggravated by virtual reality. 
The participant pool of 24 participants with homogenous demographic background limits the generalisability of the results. We hence treat the results presented below not as generalizable findings about human-automation interaction, but as preliminary evidence for the evaluation methodology proposed in this paper.

The study protocol was approved by the  Human Research Ethics Committee (HREC) of Delft University of
Technology (ID: 111053). Written informed consent was obtained from all participants prior to the experiment. At the beginning of the experiment, participants were reminded of their right to withdraw at any time without penalty. A €10 voucher was provided to each participant upon completion of the experiment. All data were anonymised and stored using unique participant identification codes.

\subsection{Driving simulator experiment}
\label{sec:driving_experiment}

The participants were asked to drive through a section of a rural road supported by driving automation; this was done repeatedly over a sequence of trials. In each trial, participants had to complete a sequence of overtaking manoeuvres in a bidirectional road, requiring them to repeatedly swerve into the opposite lane while managing oncoming traffic, while interacting with an automated driving system through a haptic steering wheel.
Participants were presented a first-person view of the driving scene using Varjo VR-3 virtual reality headset while seated in front of a  SensoWheel SD-LC steering wheel capable of providing high-fidelity haptic feedback (\cref{fig:experimental_design}). The experiment was configured in JOAN \cite{beckers2023joan} -- the framework for running experiments in the CARLA environment \cite{dosovitskiy2017carla}.
Sony WH-1000XM3 noise-cancelling headphones were used to reduce auditory distractions.

In each trial, participants drove about $1.4$ kms which took $103 \pm 5 $ seconds. With a total driving time of around 14 minutes for experimental trials,  the whole experiment for each participant (including familiarisation trials and filling in questionnaires) was completed in 90 minutes.
As the participants drove through a bidirectional rural road with straight and winding sections, the speed of the ego vehicle was fixed at 50 km/h under cruise control. Participants had no control over the accelerator or brake pedals, controlling only the steering wheel. This was done to isolate the lateral control task and ensure that participants’ behaviour primarily reflected their interaction with the steering-based haptic guidance, rather than confounding influences from longitudinal control. While driving, participants encountered right-hand traffic travelling in both directions at a constant speed of 40 km/h: nine vehicles travelling in the same direction, and five in the opposite direction~(\cref{fig:experimental_design}). Due to the speed difference, participants had to overtake the vehicles travelling in the same direction by swerving into the opposite lane. 
The vehicles were spaced in such a way that, with appropriate steering, it was possible for the participants to overtake all vehicles while maintaining sufficient safety margins to oncoming vehicles. 

The experiment included three conditions: fully manual control (baseline; participants had complete control over steering), haptic shared control (HSC), and traded control (TC). In the two latter conditions, the steering was controlled by a simple automated driving system which used a 
pure pursuit controller to follow a pre-recorded reference trajectory. 
In HSC, both the automation and the driver could apply torques on the steering wheel at the same time and thus steer the vehicle together. 
In TC, if the driver applied a torque above a threshold ($0.2\,\mathrm{N\,m}$), the automation's authority decreased to zero, giving the driver complete authority, and was only restored after the driver's torque dropped below the threshold for more than one second. If the torque applied by the human remained below the threshold for more than one second, automation torque gradually increased until it regained full authority or was intervened upon by the driver. 
Both HSC and TC were implemented based on the four design choices architecture~\cite{van2017four}. We must note that our implementations of HSC and TC do not represent the state-of-the-art and we have skipped features like adaptive reference for HSC or auditory warnings and take over requests for TC. The goal of this project was simply to evaluate two candidate control modes and thus chose two implementations that would explicitly trigger differing perceptions in drivers.

To familiarise themselves with the control modes and driving task, each participant completed four familiarisation trials: (i) manual driving without traffic, (ii) manual driving with traffic, (iii) driving with traffic under HSC, and (iv) driving with traffic under TC. The main experiment then consisted of nine trials: Trial 1 involved manual driving, while Trials 2–9 featured automated driving in either TC or HSC, with four trials per mode. The order of automated trials was randomized independently for each participant. 

To investigate participants' interaction with automation in safety-critical scenarios, we simulated a silent automation failure in every automated driving trial. Specifically, the ego vehicle was programmed to follow a trajectory leading toward a potential side collision with a motorcycle (\cref{fig:experimental_design}). Without timely driving intervention, the ego vehicle would collide with the target motorcycle. This safety-critical scenario was intentionally designed to emulate silent perception failure in SAE Level 2 partial driving automation, where drivers remain responsible for supervising the driving task and intervening when it behaves unsafely. The purpose of this design was not to evaluate the collision avoidance feature but to investigate how drivers exercise meaningful human control when automation silently fails. Only one motorcycle served as the collision target per trial; its path intersected with the automation's reference trajectory, while the other motorcycle travelled without conflict. The target motorcycle’s  position
was counterbalanced with those of other vehicles and motorcycles across trials using a Latin square design, with position distributions balanced across participants and control modes to avoid learning effects. 

Participants were instructed to remain in their lane (unless overtaking is needed), avoid collisions, and stay on the road. They were also told: ``\textit{Please note that the automation system is still in development and may falsely detect objects, potentially leading to collisions. You need to intervene when it behaves unsafely 
''}. Participant's subjective perceptions were measured after each trial using a post-trial survey (\cref{sec:post_trial_questions}). After completing all trials, a descriptive post-experiment questionnaire was used to qualitatively assess the subjective perceptions about the interaction (\cref{sec:post_experiment_questions}).

\subsection{Subjective perception questionnaires}
\label{sec:surveys}
Post-trial surveys with a rating scale were used to quantify the perception of the participants (\cref{sec:post_trial_questions}) and open ended post-experiment surveys were used to get a deeper understanding of the cues that influenced to their perception (\cref{sec:post_experiment_questions}).

\subsubsection{Post-trial subjective scores}
\label{sec:post_trial_questions}

\begin{table*}[tb]
\small
    \centering
    \begin{tabularx}{\linewidth}{>{\raggedright\arraybackslash}X c}
        \hline
        \textbf{Post-trial questions -- automated driving trials (HSC and TC)} & \textbf{Property} \\ \hline
        \textbf{A1}: I felt that I had sufficient control over the automated vehicle & Sufficient control (P3) \\
        \textbf{A2}: I felt that the automated vehicle understood my intentions during the driving task & AV understood me (P2) \\
        \textbf{A3}: I felt that I had sufficient understanding about the behaviours of automated vehicle & I understood AV (P2)\\
        \textbf{A4}: I felt that the automated vehicle and I were working together towards the same goal & Working together (P2) \\
        \textbf{A5}: I felt responsible for the driving task when I was using the automated vehicle & Responsible (P4) \\
        \textbf{A6}: I felt safe in the automated vehicle during the driving task & Safe\\
        \textbf{A7}: I trusted the automated vehicle during the driving task & Trust \\
        \hline
        \textbf{Post-trial questions -- manual driving trials} & \\ \hline
        \textbf{M1}. I felt that I had sufficient control over the vehicle. & Sufficient control~~\\
        \textbf{M2}. I felt responsible for the driving task when I was using the vehicle. & Responsible\\
        \textbf{M3}. I felt safe in the vehicle during the driving task. & Safe\\
        \hline
    \end{tabularx}
    \normalsize
    \caption{Post-trial questions that participants had to answer on a Likert scale (1 to 10) after interacting with the automated driving system through haptic shared control or traded control, and their relation to the properties of systems under meaningful human control~\cite{cavalcantesiebertMeaningfulHumanControl2022}. Questions A6 and A7 were designed to assess perceived safety and trust, concepts that are related to MHC but do not map directly onto the four MHC properties. The questions for manual driving were redacted since there was no automation involved.}
    \label{tab:post_trial_questions}
\end{table*}

Seven Likert-scale (1 to 10) questions (\cref{tab:post_trial_questions}) were asked after each HSC or TC trial to gauge how participants experienced the interaction with the driving automation (referred to as the automated vehicle (AV) in the questionnaires) and to ultimately assess the extent to which the driving
automation operated under MHC.
The questions were derived from the previously proposed actionable properties of systems under MHC \cite{cavalcantesiebertMeaningfulHumanControl2022}. 
Based on tracking and tracing conditions~\cite{santoni2018meaningful}, these properties require that:
\begin{enumerate}[topsep=0pt]
    \item [\textbf{P1}] a \emph{moral operational design domain} (moral ODD; extending the traditional notion of operational design domain towards relational and moral aspects) is clearly defined and mechanisms are in place for keeping the system in that domain;
    \item [\textbf{P2}] \emph{representations} which humans and automation have of each other, the environment, and the moral ODD,  are \textit{shared}, i.e., mutually compatible;
    \item [\textbf{P3}] humans have sufficient \emph{ability and authority to control} the outcome of the automation;
    \item [\textbf{P4}] automation \emph{actions can be explicitly linked to at least one human} who is aware of their responsibility.
\end{enumerate}

Since here we focused on the interaction between the human and the automation, rather than on the design of the automation itself,  we designed the control modes HSC and TC to have identical moral operational design domains.
Hence, in our comparison of HSC and TC we did not include any questions for Property 1.

Property 2 (\emph{shared representations}) was divided into three components: \textbf{AV understood me}, \textbf{I understood AV}, and \textbf{working together}. This division reflects Cavalcante Siebert et al.’s \cite{cavalcantesiebertMeaningfulHumanControl2022} description that shared representations between human and AI systems depend on how both agents understand each other and are able to update their representations in response to changing reasons. However, these questions do not capture aspects of Property 2 like whether the internal representations and intentions of the automated system align with those of the driver. We chose to skip these aspects in the current study, as the simplified automation in this experiment is based on a static reference trajectory that does not depend on the interaction with the human.

Property 3 (\emph{ability and authority to control}) was represented by question for \textbf{sufficient control}. Besides what is probed in this question, property 3 also pertains to the authority and ability of the relevant human to influence the system. With regards to the ability of the driver questionnaires for evaluating situational awareness~\cite{verhagen2024meaningful} and work load~\cite{sheridan1978human} would also have contributed to the assessment of property 3. However, to reduce burden on participants, we included only the question pertaining to the driver's perception of having sufficient control.

Property 4 (\emph{actions are linked to a responsible human}) was represented by the question for feeling \textbf{responsible}. It must be noted that responsibility can refer to various concepts like behaving responsibly~\cite{remyLearningResponsibilityAllocations2024}, having role responsibility, or being causally responsible~\cite{vincentMoralResponsibility2011}. In the context of property 4, Cavalcante Siebert et al. \cite{cavalcantesiebertMeaningfulHumanControl2022} note two variants, 1) the forward-link pertaining to human being aware of their moral responsibility and 2) the backward-link pertaining to the traceability of responsibility. Asking separate questions for all of these aspects of responsibility ran the risk of confusing and biasing the participants. Hence, we included a single question about perceived responsibility, and relied on the open-ended questionnaire at the end of the experiment to provide further insight into which aspects of responsibility participants thought about. 

In designing the questionnaire, we focused on specific aspects to keep questionnaires feasible for participants to complete after every trial. The questions only capture isolated aspects of the property, and are not intended as a comprehensive measure but instead an example of how certain aspects could be measured.

In addition to items directly related to the MHC properties \cite{cavalcantesiebertMeaningfulHumanControl2022}, we included questions on perceived \textbf{safety} and \textbf{trust}; as previous research has demonstrated, 
they can strongly influence evaluations of MHC based on how drivers experience interactions with driving automation
\cite{suryana2024meaningful, pengConceptualisingUserComfort2024}. Perceived safety and trust are complex constructs and should ideally be measured more comprehensively~\cite{korberTheoreticalConsiderationsDevelopment2019,nordhoffPerceivedSafetyTrust2021}. However, for similar reasons of feasibility, we simplified the measurement and used only one question for each.

As the manual driving trials did not include driving automation, only three of the above questions were applicable to these trials; different formulations were used to account for the absence of automation. Thus, answers to questions M1, M2 and M3 were used to evaluate participants' baseline perception of sufficient control, responsibility and safety, respectively (\cref{tab:post_trial_questions}).

\subsubsection{Post-experiment open-ended questions}
\label{sec:post_experiment_questions}
After all the driving trials, an open-ended post-experiment questionnaire was administered to get a deeper understanding of the perceptions of the participant and the factors that might have influenced these perceptions. 
The questions related to the same concepts mentioned in post-trial questions (\emph{sufficient control}, \emph{AV understood me}, \emph{I understood AV}, \emph{working together}, \emph{responsibility}, \emph{safety}, and \emph{trust}). Two questions per concept asked the participant to recall positive and negative experience related to that concept. For example, the questions related to \emph{sufficient control} were:\\
\\
\textbf{D1}: ``Were there any situations where you felt you \emph{had sufficient control} over the automated vehicle operation? Please describe.'' and,\\
\textbf{D2}: ``Were there any situations where you felt you \emph{did not have sufficient control} over the automated vehicle operation? Please describe.''\\
An exhaustive list of the questions is included in the supplementary material.

These items were designed to record in greater detail the overall impressions the participants had of HSC and TC, to elicit scenarios that positively or negatively affected their perception of the qualities being measured in the post-trial questionnaire,
and to compare their experiences across conditions. In particular, they addressed aspects that could not be evaluated meaningfully on a trial-by-trial basis, such as general system preference, system characteristics that affect workload, and qualitative reflections on system behaviour. This approach follows \cite{verhagen2024meaningful} in employing post-experiment reflections to evaluate MHC. In our study, such qualitative insights provide valuable context for understanding the Likert-scale scores recorded after each trial.

\subsection{Behavioural metrics and hypotheses}
\label{sec:behavioural_metrics}

\begin{figure}[tb]
    \centering
    \begin{overpic}[width=\linewidth]{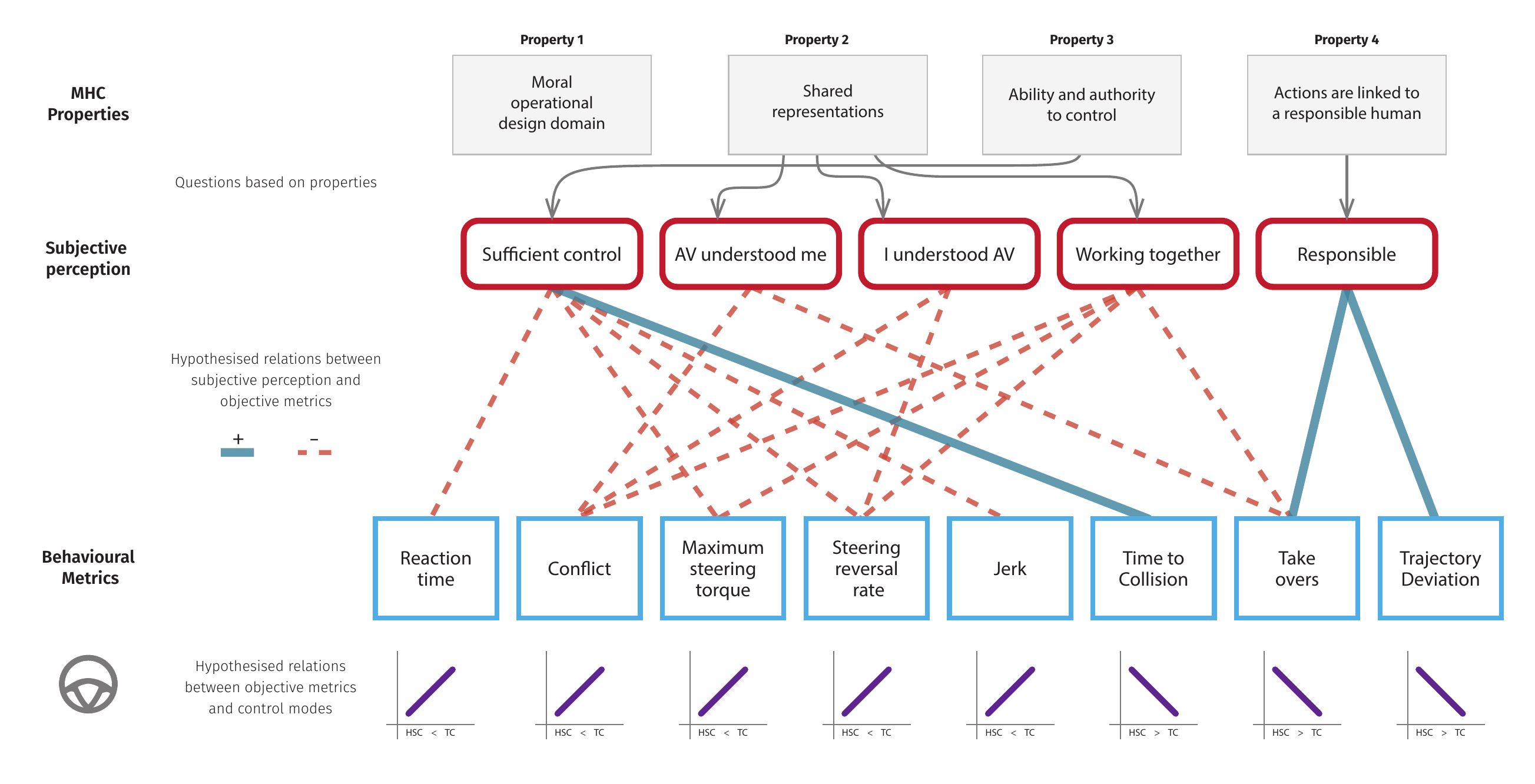}
    \put(4.75,41.5){\tiny\cite{cavalcantesiebertMeaningfulHumanControl2022}}  
  \end{overpic}
    \caption{ \textbf{Deriving hypotheses about survey questions and behavioural metrics from the properties of systems under meaningful human control (MHC)}:
    Survey questions related to MHC properties were used to evaluate subjective perception of MHC. The figure also shows the hypothesised dependence of subjective perception of MHC on behavioural metrics and control modes.}
    \label{fig:MHC_Properties}
\end{figure}

To study how subjective perceptions are related to behavioural aspects of driver-automation interaction, we formulated trial-level behavioural metrics that quantify interaction dynamics between the driver and the automation, driver's performance, and the characteristics of vehicle trajectories. We then formulated hypotheses regarding a) correlation of these metrics to certain subjective perception scores, and b) differences in these metrics between HSC and TC (\cref{fig:MHC_Properties}). Brief descriptions of metrics and the hypotheses linked to them are mentioned below; all metrics except reaction time and time to collision were calculated by aggregating the data over the whole duration of each trial.
Detailed definitions of each metric are provided in supplementary materials.

\noindent\textbf{Reaction time}: In experiment trials where a simulated silent automation failure triggered an incursion manoeuvre towards a motorcyclist, the reaction time was quantified as the time between the initiation of the automation manoeuvre and the first detectable human response. The automation initiation was identified as the first moment when the gradient of the automation torque exceeded a threshold (0.18 Nm/s), while the human response was determined based on conflict onset in HSC and authority in TC. We hypothesised that reaction times would be negatively correlated with subjective scores for item A1 (`sufficient control'). Because of the continuous nature of the interaction in HSC, we hypothesised that reaction times would be shorter for HSC than for TC.

\noindent\textbf{Average conflict in steering torques}:
In the experiment, the human driver interacted with the automation through the steering wheel. We used the conflict in steering torques --- defined as the negative signed product of human and automation applied torques that exceed a threshold (0.2 Nm) ---
to measure the disagreement between the human and the automation. In line with prior research~\cite{boink2014understanding}, we hypothesised that conflict is negatively correlated with subjective scores A2, A3, and A4 (`AV understood me', `I understood AV', and `working together', respectively). Based on the assumption that interactions with HSC will be smoother, we hypothesised that conflict will be lower for HSC than for TC.

\noindent\textbf{Maximum steering torque}: When interacting with the automation, the steering torque exerted by the participant reflects their control effort. Thus, we used maximum value of the human-applied torque during the trial as a metric of effort. Earlier studies have shown that drivers exert maximum torque when resisting lane-keeping or lane-departure assistance \cite{ercan2018predictive}.
Thus, we hypothesised that maximum steering torques are negatively correlated with scores of `sufficient control' (A1) and `working together' (A4). Due to the smoothness of the interaction in HSC, we hypothesised that maximum steering torques will be lower for HSC than in TC.

\noindent\textbf{Steering reversal rate}: 
Another established method for quantifying the control effort of a driver considers steering reversals: the greater the rate of steering reversals, the greater the control effort of the driver~\cite{mars2014analysis}. Steering reversal rate was computed as the number of steering reversals per second across a full trial, where a reversal is counted when the rate of change of steering angle is zero and the difference between two extremes exceeds 1°~\cite{markkula2006steering}. We hypothesised that the steering reversal rate is negatively correlated with the scores A1, A3, and A4 (`sufficient control', `I understood AV' and `working together', respectively). Based on the continuous nature of HSC, we hypothesised that the steering reversal rate would be lower for HSC than for TC.

\noindent\textbf{Jerk of vehicle trajectories}: The smoothness of the trajectories of the ego vehicle was quantified as the root-mean-squared jerk (longitudinal and lateral jerks combined) computed over the whole trial. We used jerk as a metric for the smoothness of the control exerted by the driver (unlike real driving, our participants did not physically experience jerk in our experiment). 
We hypothesised that trajectory jerk would be negatively correlated with the score A1 (`sufficient control'). Also, assuming that HSC would result in smoother trajectories, we hypothesised that jerk would be smaller for HSC than TC.

\noindent\textbf{Time to collision (TTC)}: In each trial where a simulated silent automation failure could lead to a collision with a motorcyclist, the minimum TTC between the ego vehicle and the motorcyclist can be used to measure the \textit{criticality} of the interaction: the lower the TTC, the more critical the interaction is. The TTC was calculated using a two-dimensional TTC algorithm~\cite{jiao2023ttc}, with the minimum taken over the duration of the failure event. 
We hypothesised that low min-TTC values (higher interaction criticality) would be associated with reduced perception of `sufficient control'; that is, we expected a positive correlation between TTC and A1. Furthermore, as the continuous engagement of the driver can reduce the criticality of the interaction in HSC, we hypothesised that TTC would be larger for HSC than for TC .

\noindent\textbf{Number of takeovers:} When a driver is interacting with the automation through TC, a takeover happens if the torque applied by the driver exceeds the pre-defined threshold which would disengage the automation. For HSC, a takeover was counted each time the conflict signal transitioned from zero to positive. The total number of such events was summed per trial. We assumed that takeovers would be triggered when the behaviour of the automation was not aligned with the intentions of the driver. Hence, we hypothesised that the number of takeovers would be negatively correlated with scores A2 (`AV understood me') and A4 (`working together'). At the same time, we hypothesised that the number of takeovers would be  
positively correlated with the score A5 (`responsible'), as the driver would have more influence on the trajectory of the ego vehicle with more takeovers.
We hypothesised that participants would be more engaged in HSC and that the number of takeovers would be greater in the case of HSC than for TC.

\noindent\textbf{Trajectory deviation}:
The deviation of the trajectory of the ego vehicle from the reference trajectory can be used to capture the extent to which the driver contributed to the behaviour of the vehicle. As an example, if the trajectory doesn't deviate at all from the reference, that would mean that the driver fully relied on automation. This was quantified as the root-mean-squared error between the actual and reference trajectories, computed over the whole duration of a trial. The reference point for each time step was taken as the nearest point on the automation's reference trajectory. We hypothesised that root-mean-squared trajectory deviation is positively correlated with the score A5 (`responsible'). Also, assuming that drivers are more engaged in HSC, we hypothesised that trajectory deviation is higher for HSC than TC.

\noindent\textbf{Overtaking time}:
To quantify how driver's preferences might deviate from the reference trajectory, we also included the overtaking time defined as the total time spent by the ego vehicle in the opposite lane during a trial.
Since the preferences of individual drivers might lead to longer or shorter overtaking times, we did not associate any hypotheses with the this metric; it was only used for exploratory analyses.

By providing objective insight into the behaviour of drivers, these metrics complement the data collected about the subjective perception of various concepts related to MHC. Detailed definitions of these metrics can be found in supplementary materials.

\subsection{Analysis}
\label{sec:analysis}

Behavioural data and subjective scores collected during the experiment were analysed quantitatively to test hypotheses (\cref{sec:AnalysisQuantitative}) and the answers to post-experiment questionnaires were analysed qualitatively to garner deeper insights into the factors affecting them (\cref{sec:AnalysisQualitative}).

\subsubsection{Quantitative analysis}
\label{sec:AnalysisQuantitative}
To analyse the effect of the control modes (HSC and TC) on the behaviour of participants, we fit linear mixed-effect models (LMMs) for each behavioural metric (\cref{sec:behavioural_metrics}): $$\text{behavioural-metric} \sim \text{control-mode} + (1|\text{participant}).$$
Here, behavioural metrics (z-scored) were the dependent variables, control mode was the independent variable with a fixed effect, and participant ID was a random intercept.

To examine how participant's perceptions during the experiment were related to their behaviour and control modes, we analysed the post-trial subjective scores (\cref{sec:post_trial_questions}) as a function of control mode and behavioural metrics.
For each subjective perception item, one LMM was fit with the
subjective score  as the dependent variable, control modes and all behavioural metrics as fixed-effect independent variables, and participant ID as a random intercept:$$\text{subjective-score} \sim \text{all-behavioural-metrics} + \text{control-mode} + (1|\text{participant}).$$ Control mode is included as a fixed effect to account for the effect of interaction type on subjective scores. The results of these LMMs were used in the confirmatory analyses of our hypotheses about the relationships between behavioural metrics and subjective scores (\cref{fig:MHC_Properties}). For the confirmatory analyses, we used the statistical significance level $\alpha = 0.05$.

In addition to testing our hypotheses, we conducted an exploratory analysis (based on the above LMMs) to identify potential relations between subjective scores and behavioural metrics that might have been overlooked in the hypotheses. The purpose of the exploratory analysis was not to make further claims, but to highlight possible relationships to investigate in future experiments. In our initial hypotheses, we had assumed that the relationships between subjective scores and behavioural metrics would be the same across control modes. However, some responses to the post-experiment questionnaire hinted that the nature of the interaction, as determined by the control mode, might also influence the relationships between subjective scores and behavioural metrics. To study whether control modes influence these fixed effects, we split the data and fit separate LMMs for each control mode. For the exploratory analysis, statistical significance of slopes was assessed at $p<0.05$. No correction for multiple comparisons was applied as we prioritised minimizing Type-II errors (false negatives) over Type-I errors (false positives). We did not formulate hypotheses about the effect of control mode on subjective scores; these relationships were therefore explored as part of the exploratory analysis only.

\subsubsection{Qualitative analysis}
\label{sec:AnalysisQualitative}

To analyse participants' responses to the post-experiment questionnaire, we categorised their free-text answers into thematic topics using a two-coder approach. Both coders reviewed participants' responses in their entirety to ensure familiarity with the dataset. An initial coding pass was conducted by one coder to identify all possible topics mentioned by participants, without restricting the scope to pre-defined categories. This exploratory, data-driven approach allowed for the inclusion of topics beyond those anticipated by the MHC framework.

Following the initial coding, similar topics were consolidated into broader thematic categories. Within each category, related ideas were organised into subtopics that together captured the range of participant experiences. This hierarchical structure enabled the preservation of nuanced perspectives while reducing redundancy and complexity in the dataset.

Once the preliminary categorisation was complete, a second coder independently reviewed the grouped topics. Both coders then developed labels and classifications for each factor, determining the most appropriate and precise naming. This independent labelling stage ensured that classification decisions were made without bias from prior discussion.

To assess coding reliability, both coders independently evaluated the presence or absence of each identified factor within participant responses. Each factor–response pair constituted one coded item for the reliability analysis. Inter-coder agreement was quantified using Cohen’s kappa \cite{cohen1960coefficient} and Gwet’s AC1 \cite{gwet2008computing}. 
See supplementary materials on osf.io for more details about the calculation of these metrics.
Discrepancies between coders were resolved through discussion until full consensus was achieved. This consensus-based, investigator-triangulation approach follows established best practices for directed content analysis \cite{braun2006using,hill2005consensual,campbell2013coding}.

\section{Results}

\subsection{Quantitative findings}
\label{sec:ResultsQuantitative}

\begin{figure}[tb]
\centering
\begin{subfigure}[c]{\linewidth}
    \centering
    \begin{subfigure}[c]{\linewidth}
    \begin{overpic}[width=\linewidth]{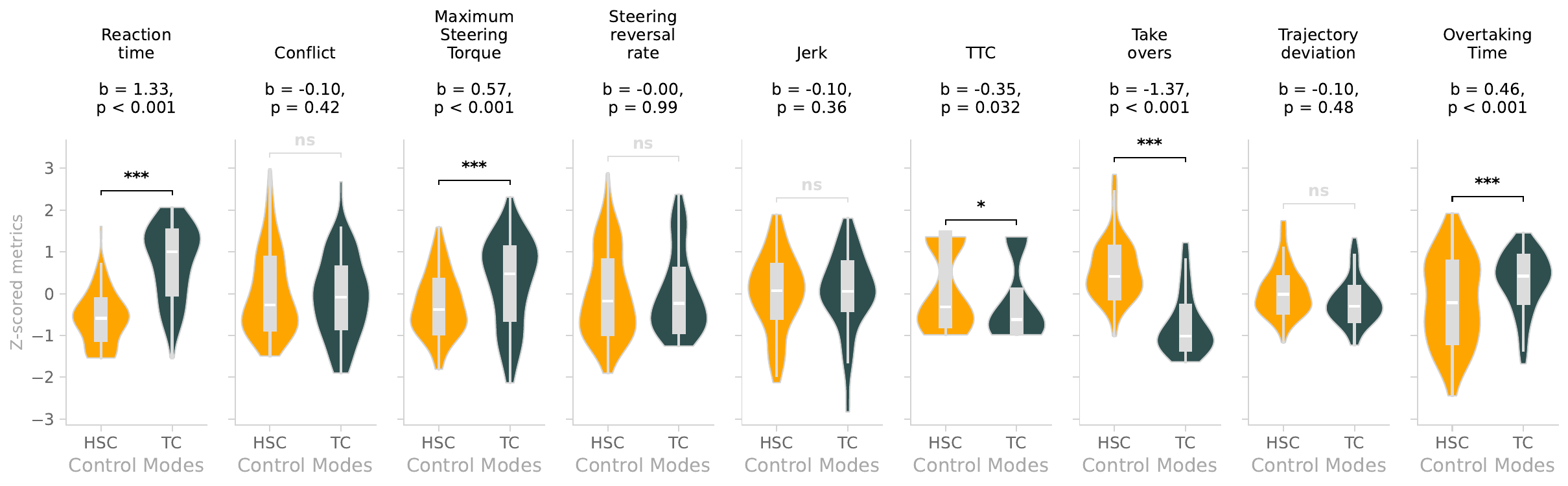}
    \put(0,1){(a)}
    \end{overpic}
    \end{subfigure}
\label{fig:Mixed_Model_metric_v_control_mode}
\end{subfigure}
\begin{subfigure}[c]{\linewidth}
    \centering
    \begin{overpic}[width=\linewidth]{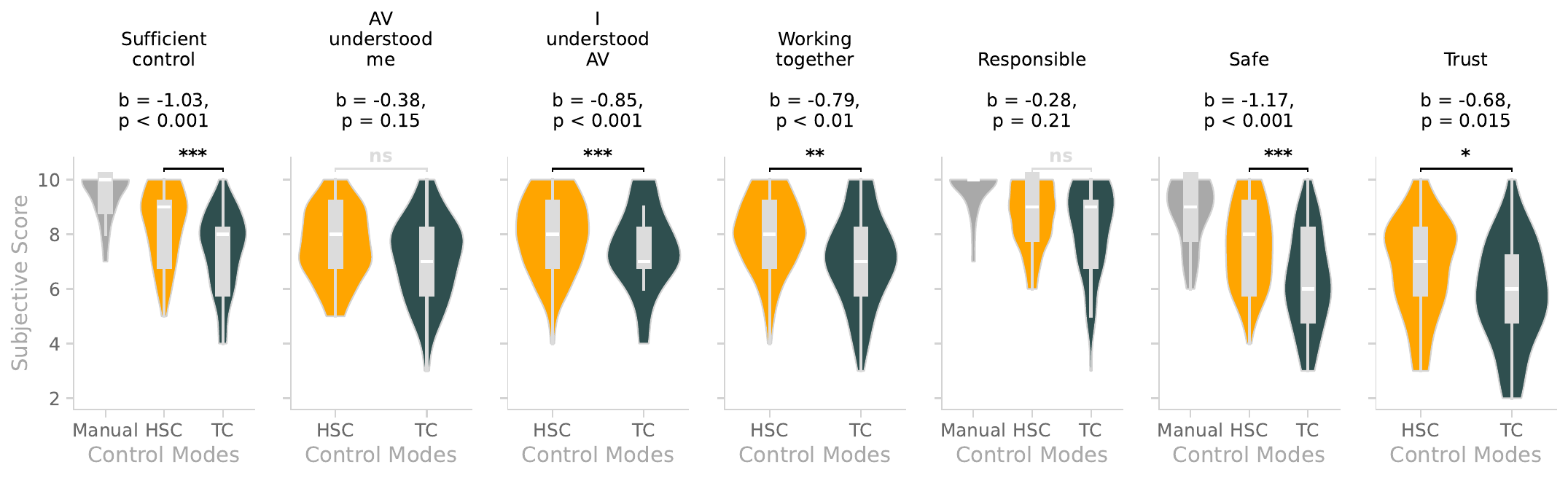}
    \put(0,1){(b)}
    \end{overpic}
    \label{fig:Mixed_Model_Q_v_control_mode}
\end{subfigure}
\caption{\textbf{The effects of control mode on behavioural metrics (confirmatory analysis) (a) and subjective perception scores (exploratory analysis) (b)}.
Statistical comparisons in panel (a) are based on confirmatory analyses using linear mixed-effect models  (Section~\ref{sec:AnalysisQuantitative}; \cref{tab:Hypotheses_metrics_vs_control_modes}). Comparisons in panel (b) are exploratory and indicated for reference only. Statistical tests are based on LMMs used for testing the relationships between behavioural metrics and subjective scores (Section~\ref{sec:AnalysisQuantitative}) which also included control mode as the fixed effect; 
significance codes: \emph{$^{***}$~0.001 $^{**}$~0.01 $^*$~0.05}. For visualization purposes, outliers (z-scores $> 3$) were excluded in panel (a).
}
\label{fig:resultsQuantitativeControlMode}
\end{figure}

\begin{figure}[tb]
\centering
\includegraphics[width=1\linewidth]{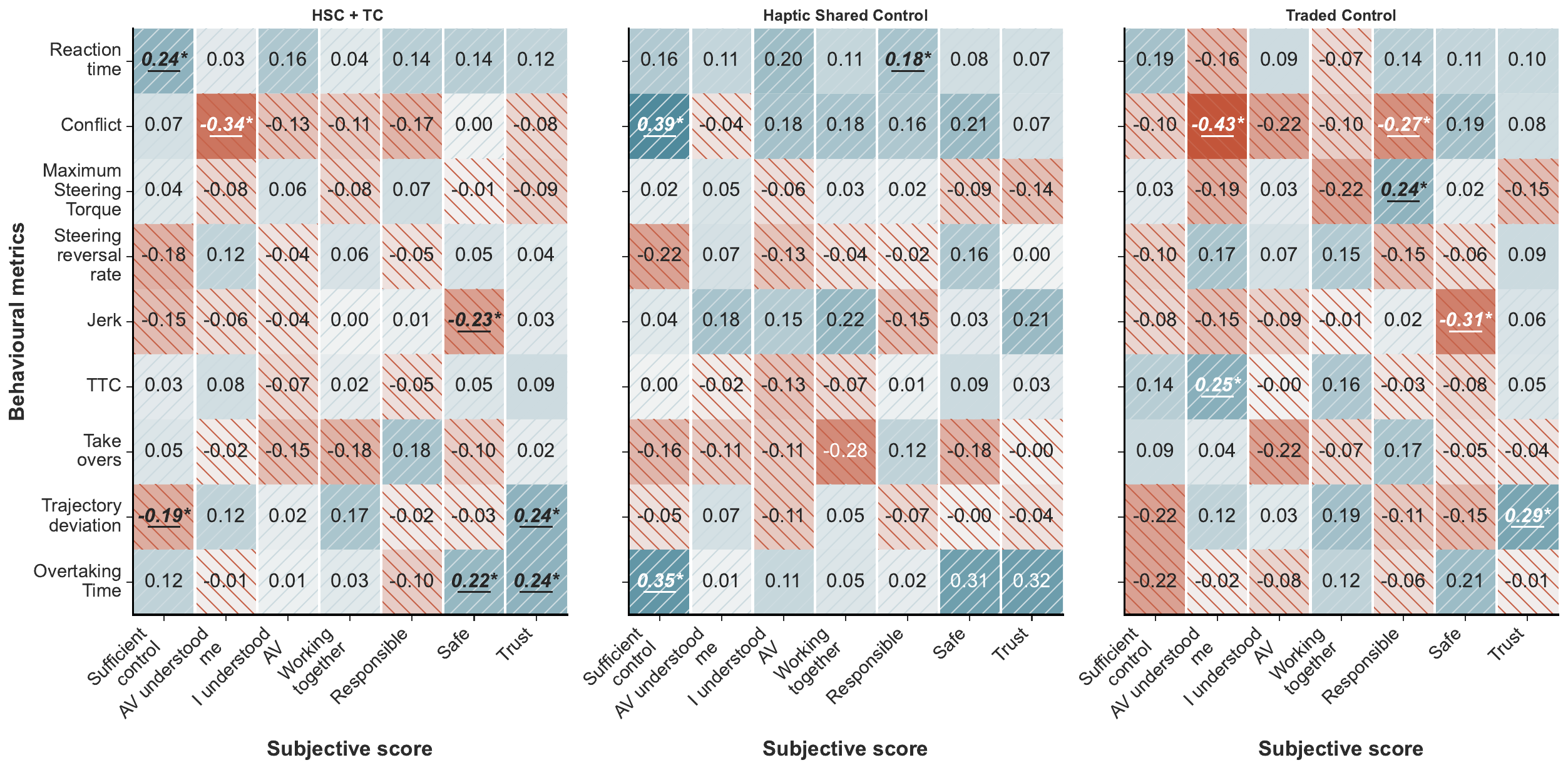}
\caption{\textbf{Exploratory analysis --- Subjective scores vs behavioural metrics:}  The slope coefficients for each metric obtained from the linear mixed-model regressions with the formula: $\text{subjective-score} \sim \text{all-behavioural-metrics} + \text{control-mode} + (1|\text{participant})$; p<0.05 is indicated by $^*$.
}
\label{fig:Mixed_Model_Q_v_metric}
\end{figure}

The quantitative results are reported in
\cref{tab:Hypotheses_metrics_vs_control_modes,tab:Hypotheses_perception_vs_metrics,fig:resultsQuantitativeControlMode,fig:Mixed_Model_Q_v_metric}.
The confirmatory analysis tested two sets of hypotheses: the effect of control mode on each behavioural metric (\cref{fig:resultsQuantitativeControlMode,tab:Hypotheses_metrics_vs_control_modes}), and the relationships between behavioural metrics and subjective scores (\cref{tab:Hypotheses_perception_vs_metrics}). The exploratory analysis examined additional relationships between behavioural metrics and subjective scores that were not part of the original hypotheses (Fig.~\ref{fig:Mixed_Model_Q_v_metric}). For each behavioural metric below, we first report the confirmatory relationship with control mode, then with subjective score, and finally any additional findings.

\begin{table}[tb]
	\centering
    \small
\begin{tabular}{>{\raggedright\arraybackslash}m{3.5cm}>{\centering\arraybackslash}m{3cm}>{\centering\arraybackslash}m{2cm}>{\centering\arraybackslash}m{2cm}>{\centering\arraybackslash}m{2cm}>{\centering\arraybackslash}m{2cm}}
	\hline
	\textbf{Behavioural Metric} & \textbf{Hypothesised relation} & \textbf{Observed relation} & \textbf{Slope $\beta$} &  \textbf{p-value} & \textbf{Hypothesis accepted} \\
	\hline
	Reaction time           & HSC < TC & < & \textit{~1.33} & \textit{< 0.001$^{***}$} & \checkmark \\
	Conflict                & HSC < TC &   & -0.10~ & ~0.42~ & \\
	Maximum steering torque & HSC < TC & < & \textit{~0.57} & \textit{< 0.001$^{***}$} & \checkmark \\
	Steering reversal rate  & HSC < TC &   & -0.00~ & ~0.99~ & \\
	Jerk                    & HSC < TC &   & -0.10~ & ~0.36~ & \\
    TTC                     & HSC > TC & > & \textit{-0.35}~ & ~\textit{0.03~$^*$}~ &  \checkmark \\
	Takeovers              & HSC > TC & > & \textit{-1.37} & \textit{< 0.001$^{***}$} & \checkmark \\
	Trajectory deviation    & HSC > TC &   & -0.10~ & ~0.48~ & \\
	\hline
\end{tabular}
\normalsize
    \caption{\textbf{Confirmatory analysis -- behavioural metrics vs control modes:} Hypothesised relations and results of statistical tests based on LMMs: $\text{behavioural-metric} \sim \text{control-mode} + (1|\text{participant})$ (Section~\ref{sec:AnalysisQuantitative}); significance codes: \emph{$^{***}$~0.001 $^{**}$~0.01 $^*$~0.05}. $\beta$ is TC relative to HSC.
    }
	\label{tab:Hypotheses_metrics_vs_control_modes}
\end{table}

\begin{table}[htb]
\small
    \centering
\begin{tabular}{>{\raggedright\arraybackslash}m{3cm}>{\raggedright\arraybackslash}m{3.5cm}>{\centering\arraybackslash}m{2cm}>{\centering\arraybackslash}m{1.5cm}>{\centering\arraybackslash}m{2cm}>{\centering\arraybackslash}m{1.5cm}>{\centering\arraybackslash}m{2cm}}
	\hline
	\textbf{Subjective score} & \textbf{Behavioural metric} & \textbf{Hypothesised relation} & \textbf{Slope $\beta$} &  \textbf{p-value} 
    & \textbf{Hypothesis accepted} \\
	\hline
	Sufficient control  & Reaction time            & - & \textit{~0.238} & \textit{0.014 $^*$}\\
	                    & Maximum steering torque  & - & ~0.038~ & 0.666~\\
                    	  & Steering reversal rate   & - & -0.178~ & 0.086~\\
                    	  & Jerk                     & - & -0.148~ & 0.083~\\
                        & TTC                      & + & ~0.033~ & 0.621~\\
	\hline
	AV understood me    & Conflict                 & - & \textit{-0.344} & \textit{0.001$^{***}$}&\checkmark\\
	                    & Takeovers               & - & -0.019~ & 0.880\\
	\hline
	I understood AV     & Conflict                 & - & -0.130~ & 0.203\\
	                    & Steering reversal rate   & - & -0.038~ & 0.720\\
	\hline
	Working together    & Conflict                 & - & -0.110~ & 0.307\\
                    	  & Maximum steering torque  & - & -0.080~ & 0.402\\
                    	  & Steering reversal rate   & - & ~0.061~ & 0.565\\
                    	  & Takeovers               & - & -0.176~ & 0.167\\
	\hline
	Responsible         & Takeovers               & + & ~0.182~ & 0.900~\\
                        & Trajectory deviation     & + & -0.022~ & 0.769\\
	\hline
\end{tabular}
\normalsize
    \caption{\textbf{Confirmatory analysis --- subjective scores vs behavioural metrics:} 
    Hypothesised relations and results of statistical tests based on LMMs $\text{subjective-score} \sim \text{behavioural-metric} + \text{control\_mode} + (1|\text{participant})$ (Section~\ref{sec:AnalysisQuantitative}); significance codes: \emph{$^{***}$~0.001 $^{**}$~0.01 $^*$~0.05}.
    `+' indicates that the perception score was hypothesized to be positively correlated with the corresponding behavioural metric; `-' indicates hypothesised negative correlation.
    }
    \label{tab:Hypotheses_perception_vs_metrics}
\end{table}

\newpage
\mysub{Reaction time}
In accordance with our hypothesis, reaction times in simulated automation failures were significantly lower for HSC than for TC ($\beta=1.33$, $p<0.001$). However, contrary to our hypothesis, reaction times were positively correlated with the score for `sufficient control' ($\beta=0.24$, $p=0.01$). The exploratory analysis also suggested a positive correlation between reaction time and the score of `responsible' for HSC ($\beta=0.18$, $p=0.04$).  

\mysub{Conflict in steering torques}
There was no evidence for the hypothesized difference in conflict in steering torques between HSC and TC ($\beta=-0.1$, $p=0.42$). 
At the same time, the hypothesis that conflict in steering is negatively correlated with the score for `AV understood me' was confirmed ($\beta=-0.34$, $p=0.001$). The exploratory analysis suggested that conflict might be positively correlated with the score of `sufficient control' for HSC ($\beta=0.39$, $p=0.01$) and negatively correlated with the score of `responsible' for TC ($\beta=-0.27$, $p=0.03$).  

\mysub{Maximum steering torque}
Consistent with our hypothesis, maximum steering torque was lower for HSC than for TC ($\beta=0.57$, $p<0.001$). The exploratory analysis indicated a possible positive correlation between maximum steering torque and the score of `responsible' for TC ($\beta=0.235$, $p=0.04$). 

\mysub{Steering reversal rate}
There was no evidence for relationships between steering reversal rate and control mode or subjective scores.

\mysub{Jerk of vehicle trajectories}
The data showed no evidence for differences in the jerk of vehicle trajectories between HSC and TC. None of the hypothesised correlations of jerk with subjective scores were substantiated in the confirmatory analysis. The exploratory analysis showed that trajectory jerk could be negatively correlated with the score of `safe' ($\beta=-0.22$, $p=0.01$), especially for TC ($\beta=-0.314$, $p=0.01$).

\mysub{Time to collision}
The minimum TTC during silent automation failure was longer for HSC than for TC ($\beta=-0.35$, $p=0.03$). There was no evidence for the hypothesised correlation between TTC and the score of `sufficient control' either. However, the exploratory analysis suggested a positive correlation between minimum TTC and the score for `AV understood me' for TC ($\beta=-0.253$, $p=0.03$). 

\mysub{Number of takeovers}
As per our hypothesis, the number of takeovers was larger for HSC than for TC ($\beta=-1.37$, $p<0.001$). No evidence was however found for correlations between the number of takeovers and any of the subjective scores.

\mysub{Trajectory deviation}
No evidence was found for difference in deviation of the vehicle trajectory from pre-recorded reference between HSC and TC. The confirmatory analysis did not yield evidence for any of the hypothesised correlations of trajectory deviation with subjective scores. The exploratory analysis indicated that trajectory deviation can be negatively correlated with the subjective score for `sufficient control' ($\beta=-0.19$, $p=0.02$) and positively correlated with the score for `trust' ($\beta=0.24$, $p=0.01$), especially for TC ($\beta=0.29$, $p=0.05$).

\mysub{Overtaking time}
Overtaking time was not part of our hypotheses; exploratory analyses suggested that overtaking time was larger for TC than for HSC ($\beta=0.46$, $p<0.001$). Overtaking time might be positively correlated with subjective score for 'safety' ($\beta=0.22$, $p=0.04$) and 'trust' ($\beta=0.24$, $p=0.05$). Additionally, for HSC, we found potential positive correlation between overtaking time and the score for `sufficient control' ($\beta=0.35$, $p=0.01$).

\subsection{Qualitative findings}
\label{sec:ResultsQualitative}

We performed a topic analysis of participants' responses to the post-experiment questionnaire to gain deeper insight into the factors that influenced their subjective experience by categorising those experiences into thematic topics. As part of the qualitative analysis, inter-coder agreement was assessed on a set of 57 coded factors. The observed agreement between coders was 66.67\%, with Cohen’s kappa \cite{cohen1960coefficient} of –0.1993 and Gwet’s AC1 \cite{gwet2008computing} of 0.5385. 

The negative kappa value was attributable to the absence of cases where both coders labeled a factor as absent  (see contingency table in supplementary materials for details), 
which can produce prevalence bias in kappa calculations \cite{feinstein1990high}. The agreement between both coders was on the presence of a factor in 38 cases, with disagreement in 19 cases (nine coded as present by Coder 1 only, ten coded as present by Coder 2 only), and no instances where both coded the factor as absent. Given these conditions, Gwet’s AC1 was taken as the more robust measure of agreement.

Full consensus was reached after discussions between coders consolidated
overlapping factors and resolved disagreements by jointly evaluating the arguments presented by each coder.
This process resulted in 47 final factors that either positively or negatively affected perceptions related to MHC. 
These factors along with remarks for control modes are summarised in \cref{tab:QA_summary}.
We classified the factors into three categories:
\textbf{(1) attitudinal factors} \attitudinal relating to mental states of participants,
\textbf{(2) interaction factors} \interaction relating to human-AV interaction through the steering wheel,
and
\textbf{(3) trajectory factors} \trajectory relating to the motion of vehicles.
We follow this categorisation when describing the findings for each type of perception. Participant quotes are referenced by anonymised IDs (e.g., ID61, ID90); these are randomised identifiers and do not reflect the sequential numbering of the 24 participants.

\mysub{Sufficient Control}
Within the epistemic dimension, participants reported that HSC gave them a greater \textit{sense of control}, describing that in most cases they felt they had sufficient control over the AV’s operation (ID61, ID90). Regarding trajectory-related factors, one participant noted inability to \textit{influence vehicle speed}
as a factor that reduced their perception of having sufficient control (ID58). In terms of interaction, the \textit{ease of overriding} the AV’s actions and \textit{continuous access to the steering wheel} were identified as factors that enhanced the perception of control (ID17, ID58, ID172). Conversely, \textit{steering forces against driver intention} in both TC and HSC (ID106, ID203) created a sense that control was being taken away rather than shared. Furthermore, \textit{readaptation effort after abrupt takeovers} (ID61, ID90, ID172, ID242),  \textit{need for continuous vigilance} (ID19), and \textit{invasive forces in TC}
(ID203) were cited as reducing the sense of control.

\newpage
\begin{longtable}[t]{%
	>{\raggedright\arraybackslash}p{(\columnwidth - 6\tabcolsep) * \real{0.05}}@{\hspace{1em}}
	>{\raggedright\arraybackslash}p{(\columnwidth - 6\tabcolsep) * \real{0.35}}@{\hspace{1em}}
	>{\raggedright\arraybackslash}p{(\columnwidth - 6\tabcolsep) * \real{0.35}}@{\hspace{1em}}
	>{\raggedright\arraybackslash}p{(\columnwidth - 6\tabcolsep) * \real{0.2}}}
\\
\toprule
\begin{minipage}[b]{\linewidth}\raggedright
\emph{}
\end{minipage} & \begin{minipage}[b]{\linewidth}\centering
\textbf{Positive}
\end{minipage} & \begin{minipage}[b]{\linewidth}\centering
\textbf{Negative}
\end{minipage} & \begin{minipage}[b]{\linewidth}\centering
\textbf{Control mode}
\end{minipage} \\
\midrule
\endhead
\multicolumn{4}{l}{\textbf{\textit{Sufficient Control}}}\\ \midrule
& & \attitudinal Need for continuous vigilance & \multirow{2}{(\columnwidth - 6\tabcolsep) * \real{0.2}}{\attitudinal  Greater sense of control for HSC}\\
& \interaction Continuous access to control (HSC) & \interaction Abruptness of takeovers (TC) & \\
& \interaction Override capability & \interaction Force against driver intention & \multirow{2}{(\columnwidth - 6\tabcolsep) * \real{0.2}}{\interaction Forces in TC more invasive}\\
& & \interaction Control overshoots & \\
& & \interaction Disruptions in fluency & \\
\midrule
\multicolumn{4}{l}{\textbf{\textit{AV understood me}}}\\ \midrule
& \attitudinal Anticipation of intention & \attitudinal Mismatched anticipation of intention & \\
& \interaction Subtle torque feedback & \attitudinal Lack of safety (TC) & \\
& \interaction Low steering resistance & \interaction High steering resistance & \\
& \interaction Override capability (HSC) & & \\
& \trajectory Timing & \trajectory Mismatched timing & \\
& \trajectory Smooth lane alignment & & \\
\midrule
\multicolumn{4}{l}{\textbf{\textit{I understood AV}}}\\ \midrule
& \attitudinal Transparency of AV intention & \attitudinal Lack of transparency of AV intention
& \\
& \interaction Clear takeovers & \attitudinal Lack of safety & \\
& \interaction Subtle torque feedback & \interaction Stronger torque feedback & \\
& & \interaction Unnecessary steering jerks & \\
& & \trajectory Erratic trajectories & \\
& & \trajectory Mismatched timing & \\
\midrule
\multicolumn{4}{l}{\textbf{\textit{Working together}}}\\ \midrule
& \attitudinal Effortless coordination (HSC) & & \multirow{2}{(\columnwidth - 6\tabcolsep) * \real{0.2}}{\attitudinal Drivers preferring autonomy, preferred HSC} \\
& \attitudinal Shared direction (HSC) & & \\
& \attitudinal Safety & \attitudinal Lack of safety & \\
& \interaction Subtle AV corrections matching driver intent & \interaction Driver corrections due to mismatched intent & \\
\midrule
\multicolumn{4}{l}{\textbf{\textit{Responsible}}}\\ \midrule
& \attitudinal~Need for supervision (TC), \&~Expectation to correct (HSC) & \attitudinal Redundancy of human input & \multirow{2}{(\columnwidth - 6\tabcolsep) * \real{0.2}}{\attitudinal Control mode defined responsibility}\\
& \attitudinal Different driving style & \attitudinal Shared-intent moments (TC) & \\
& \attitudinal Safety-critical situations & \attitudinal Monotonous driving & \\
& \attitudinal Novelty and untrustworthiness & & \\
& \attitudinal Accountability (moral, legal) & \interaction Physically disengaged (TC) &  \\
& \interaction Low steering resistance & \interaction Force against driver intention (HSC) & \\
\bottomrule
\caption{\textbf{Factors affecting the subjective perception:}
The qualitative analysis of post-experiment answers revealed factors that participants believed to \emph{positively} and \emph{negatively} affect their subjective perception. Remarks pertaining to the nature of the \emph{control modes} were also summarised. These factors were classified into three categories as 1) those pertaining to \emph{attitudinal} \attitudinal factors, 2) those pertaining to the \emph{interaction} \interaction between the human and the automation, and 3) those pertaining to the  \emph{trajectory} \trajectory of the vehicle. Similar themes repeated across columns are mentioned in the same row whenever possible. Control modes in parentheses, (HSC) and (TC), indicate factors that were associated with a specific control mode.}
\normalsize
\label{tab:QA_summary}
\end{longtable}

\mysub{AV understood me}
Perception of whether the AV understood the driver often depended on its ability to anticipate intentions accurately and act safely. In terms of attitudinal factors, TC was sometimes linked to \textit{unsafe lane changes near adjacent motorcycles} (ID106, ID172). In terms of trajectory-related factors, participants reported that  \textit{accurate anticipation of manoeuvre intentions}, such as overtaking or aligning within the lane, were associated with participants feeling that the AV understood them (ID167, ID172). Conversely, \textit{mismatched anticipation}, such as returning to the lane too early and creating a potential crash situation, made them feel misunderstood (ID229). Finally, from the perspective of human–automation interaction, \textit{low steering resistance} (ID17) and \textit{immediate handover upon intervention} (ID61) were taken as behaviours that aligned with the perception that the system understood the driver. Conversely, \textit{strong steering resistance against driver preferences} (ID264) reduced understanding.

\mysub{I understood AV}
Understanding of the AV’s behaviour was often shaped by the predictability of its actions. In terms of attitudinal factors, participants noted \textit{that familiarity with how HSC and TC operation}, gained during the experiment, made them understand the AV’s intentions (ID17, ID58). As for trajectory-related factors, several negative experiences were reported, including \textit{erratic deviations from the reference trajectory}, which could occur when driver inputs conflicted with the automation's path-following controller 
(ID302),\textit{unsafe manoeuvres} that could lead to a crash (ID454), and \textit{mismatched manoeuvre timing} that were either too early or too late (ID137, ID203). These actions influenced participants’ confidence in anticipating the AV’s next move. Regarding human–automation interaction, \textit{subtle torque feedback} was perceived as a helpful indicator for understanding the AV’s intentions (ID58, ID61), while \textit{overly strong torque feedback} (ID172) or  \textit{unnecessary steering jerks}  (ID17) were seen as intrusive factors.

\mysub{Working together}
Among attitudinal factors, the perceived \textit{sense of cooperation between the driver and the AV} was associated with HSC (ID172). In trajectory-related factors, \textit{safe manoeuvres} such as avoiding a collision
with the vehicle in front (ID227) were viewed as signs of effective teamwork, whereas \textit{unsafe manoeuvres} that could potentially lead to a crash with other vehicles undermined this perception (ID224). For human–automation interaction, \textit{subtle steering corrections} provided by the AV (ID224), \textit{minimal override pressure} in HSC (ID17), and \textit{sense of shared execution} (ID40) reinforced the feeling that participants and the AV were working together. However, \textit{driver corrections due to mismatched intent} were seen as undermining the feeling of working together with the automation systems (ID58).

\mysub{Responsibility}
Participants’ awareness of responsibility was shaped by how they understood their role, the driving situation, and the nature of their interaction with the AV. In terms of attitudinal factors, \textit{recognition of operational role} influences awareness of responsibility to supervise the automation as supervisor in TC and as driver in HSC (ID61), both of which positively influenced their awareness of responsibility. Others acknowledged \textit{knowing they were the responsible party} even when they were not actively controlling the vehicle (ID172). Counterintuitively, \textit{low level of trust in the system} positively influenced participants’ perception of responsibility, with some attributing this to their first time driving such a system (ID203). However, \textit{perception of partial responsibility}, in which the participant could intervene with the AV but also recognised that the AV could complete the manoeuvre without their input, led to ambiguity that reduced their feeling of responsibility (ID17). As for trajectory-related factors, \textit{safety-critical situations} such as overtaking (ID430), led to a stronger perception of responsibility, while \textit{monotonous driving} on a straight road (ID278) and \textit{low-traffic situations} (ID203) were associated with feeling less responsible. Finally, for human-automation interaction, both HSC and TC were reported as causing negative perceptions of responsibility. In TC, \textit{passive disengagement from the steering wheel} reduced the sense of responsibility (ID90). In HSC, \textit{feeling forced into certain moves} made them feel less responsible (ID17).

\section{Discussion}
This study set out to examine how meaningful human control can be evaluated in the context of partially automated driving, with a particular focus on linking objective behavioural measures to drivers’ subjective experience. Rather than treating control as a purely technical property of the system, we approach meaningful human control as something that emerges through human–automation interaction and is reflected in both observable behaviour and perceived experience. In doing so, we introduced a mixed-methods approach that enables these different perspectives to be analysed in relation to one another. The following discussion synthesises our quantitative and qualitative findings and reflects on how the MHC approach can inform the understanding, design, and evaluation of human–automation interaction.

\subsection{Factors related to subjective perception of control and responsibility}
\label{sec:Discussion-subjectivePerception}

\subsubsection{Sufficient Control (MHC Property 3)}
None of our hypotheses regarding correlation of behavioural metrics with the perception of `sufficient control' were accepted in the confirmatory analysis.
Interestingly, we found a positive association between the reaction time in silent automation failures and the subjective score for 'sufficient control'; this contradicted our hypothesis, as we originally assumed that faster reaction times would be associated with stronger perception of control. However, stronger perception of control might have made participants more complacent causing them to react late as found in previous studies \cite{payre2016fully, dixit2016autonomous}. This would mean that participants would have greater reaction times for the control mode for which they had greater sense of control. However, when comparing the control modes, HSC has a higher score for `sufficient control' and a lower reaction time than TC, and thereby contradicts the complacency explanation. Another plausible explanation might lie in the dynamics of the interaction: when participants react early, they might have to make smaller corrections than when they react late, which might lead to a greater sense of control in cases where they react late. 

This presents a potential contradiction where humans might experience a greater sense of agency for late reactions of greater magnitude than smaller reactions of small magnitude.
Future research can explore this dynamic with regard to other factors associated with the `sense of agency' such as sensory attenuation (where agents perceive their own actions to be of a smaller magnitude) and intentional binding (where agents perceive that the time interval between them doing some action and its outcome is smaller than the actual time elapsed) \cite{wenSenseAgencyPerception2022, berberianAutomationTechnologySense2012}.
Sensory attenuation might cause humans to feel that they are not doing much when they are performing small actions when reacting fast \cite{baysAttenuationSelfGeneratedTactile2006}.
Intentional binding might be caused by the additional cognitive load triggered when humans are being active  \cite{wenkeHowVoluntaryActions2009}, which might also explain why reaction times are higher when they are active. 

\subsubsection{AV understood me (MHC Property 2)}
In accordance with our hypothesis, the score for `AV understood me' was negatively correlated with conflict in steering torques. This is consistent with participants’ qualitative reports that \emph{force against their intention} or \emph{invasive steering torque} undermined the feeling that the AV understood them. In other words, when drivers experienced strong resistance from the automation, they interpreted it as misalignment of intent. Furthermore, the exploratory analysis for TC alone revealed a possible positive correlation between time to collision (TTC) and the score for `AV understood me'. 
This tentative correlation with TTC, along with the qualitative responses mentioning the negative impact of `mismatched intentions' and `mismatched timing', hints that the perception of `AV understood me' is sensitive to event-level responsiveness of the automation system and that it might vary within trials.
Thus, instead of aggregating perception scores and behavioural metrics per trial, future work could explore how perception varies in real time with respect to behavioural signals of finer temporal resolution.

Also, it must be noted that in our experiment, the AV never truly `understands' the driver. The controller operates with fixed parameters that do not adapt to individual drivers during interaction. This means that the perception of `AV understood me' reflects how well the design of the automation aligned with driver expectations, rather than responsiveness to driver intent. Consequently, our findings should be interpreted as evidence that certain controller characteristics, such as low conflict in steering torques, can give rise to the feeling of being understood, even in the absence of real adaptive behaviour. This has implications for future AV design: fostering the perception of mutual understanding may not require intelligence or adaptivity per se, but rather careful tuning of interaction parameters.

\subsubsection{I understood AV (MHC Property 2)}
The lack of evidence for relationships between the subjective score for `I understood AV' and the behavioural metrics might suggest that the metrics employed did not capture the interaction properties most relevant to this perception. 
The possible difference in scores between HSC and TC suggested by the exploratory analysis could in turn indicate that this perception is sensitive to control mode characteristics.
This motivates future work on identifying behavioural metrics that are more sensitive to the aspects of human--automation interaction that underlie this perception.

Qualitative reports indicated that mismatched manoeuvre timing, 
intrusive torque feedback, and erratic or unsafe manoeuvres 
undermined perceived understanding (see Section~\ref{sec:ResultsQualitative}). Taken together, the results indicate that perceived understanding may be less
about aggregate measures of steering input computed over an 
entire trial, and more about specific moments of alignment or misalignment between the AV’s actions and the driver’s expectations. Future analyses should therefore 
investigate event-based indicators (e.g., timing of manoeuvres, mismatched torque events) over global metrics (e.g., root-mean-squared trajectory deviation), which are too coarse to capture these dynamics.

\subsubsection{Working together (MHC Property 2)}
Exploratory analyses suggested that the scores for `working together' differed significantly between HSC and TC; if confirmed by future studies, this would align with qualitative reports where participants described HSC as fostering a greater sense of cooperation. However, similar to `I understood AV', no significant correlations were found between the scores for `working together' and any of the behavioural metrics.
Thus, the behavioural metrics examined in this study do not capture perceptions of `working together' with the automation; future research should explore the use of other metrics for this purpose.

These results suggest that the perception of working together might not be explained simply by metrics aggregated over trials.
Instead, qualitative reports suggest that the perception of 'working together' was sensitive to the interactional qualities for specific moments or portions of the interaction.
For example, whether corrections feel subtle or forceful, whether overrides are smooth or resisted, and whether driver and AV actions are aligned in timing and intent.
Relatedly, \cite{mars2014analysis} emphasized the complementary perspective of conflict in cooperation, noting that to capture such moments of opposition (the ``inverse'' of working together), it is more informative to compute transient torque variations around specific events rather than relying solely on global measures aggregated across entire trials.

\subsubsection{Responsibility (MHC Property 4)}
The hypothesised relationships between the score for `responsible' and behavioural metrics (number of takeovers and trajectory deviation from the reference) were not confirmed by the data. That said, the exploratory analysis for HSC indicated a potential positive correlation between the score for `responsible' and reaction time. Further, for the case of TC, it suggested that the score for `responsible' might be negatively correlated with conflict and positively correlated with maximum steering torque. 
These exploratory findings may suggest that larger or later interventions are more salient to drivers, though this interpretation remains tentative. Qualitative findings similarly linked responsibility to situations requiring supervision or correction. However, these effects must be considered alongside safety, as increasing the need for intervention may conflict with safe system behaviour. 

Overall, these results hint that the perception of being `responsible' could be diminished in cases of only subtle actions by the driver. Together with earlier findings that outside observers might wrongly blame the drivers of partially automated vehicles for accidents~\cite{beckersDriversPartiallyAutomated2022}, this warrants the need for objective metrics for ascribing responsibility in traffic interactions~\cite{georgeFeasibleActionSpace2025,shalev-shwartzVisionZeroProvable2019,remyLearningResponsibilityAllocations2024}. 

\subsubsection{Safety and trust}
 
Perceived safety~\cite{pengConceptualisingUserComfort2024, chenEvaluatingSafetyEfficiency2024, papadimitriouCommonEthicalSafe2022} and trust~\cite{payre2016fully, dixit2016autonomous, nordhoff2023mis, molnar2018understanding} have been extensively studied in the context of human-AV interactions. 
Even though perceived safety and trust were not central to our investigation of driver's MHC perceptions, we opted to include them to explore potential relationships between these concepts and the behavioural metrics that are potential MHC correlates.

Our exploratory analysis suggested that the scores for `safe' might be negatively correlated with the jerk of vehicle trajectories and positively correlated with overtaking time. This would be aligned with the earlier work which hypothesized that smoother vehicle trajectories would improve the sense of safety \cite{pengConceptualisingUserComfort2024}. 
The exploratory analysis also suggested positive associations between scores for `trust' and trajectory deviation and overtaking time. However, these findings should be interpreted with caution, as they may reflect context-specific factors rather than general behavioural patterns. In particular, in our experimental scenario overtaking time may be linked to perceived safety, which could in turn influence trust. As such, these results may not generalise beyond our specific setting.

\subsection{Empirical comparison of two control modes in relation to MHC}
\label{sec:Discussion-comparison-control-modes}
In this study, we illustrated the proposed MHC assessment methodology by evaluating drivers' interaction with automation through two control modes well-established in the literature --- haptic shared control~(HSC) and traded control~(TC). It must be noted that the following discussion only pertains to control modes HSC and TC as we have implemented them, and that these might not necessarily generalise to all forms of HSC and TC, for instance, implementations that issue takeover requests as auditory cues. 

Exploratory analyses of subjective scores collected from post-trial surveys indicate that HSC might lead to better perceptions of MHC compared to TC~(\cref{fig:Mixed_Model_Q_v_control_mode}).
In our implementation, the driving automation (i.e., its control algorithm) did not adapt to individual participants; this might have contributed to the observed lack of difference in the scores of \emph{`AV understood me'} for the two control modes.
The exploratory analysis pointed to participants reporting higher scores for \emph{`sufficient control'}, \emph{`I understood the AV'}, and \emph{`working together'} for HSC.
In their qualitative answers, participants mentioned how the forces in TC were more invasive and that HSC offered greater sense of control, especially for drivers who preferred being autonomous. 
Together, these map onto MHC properties 2 and 3, indicating better-aligned representations and improved control ability for HSC. Furthermore, \emph{`perceived safety'} and \emph{`trust'} were also rated higher for HSC compared to TC.

In terms of behavioural outcomes, participants showed faster reaction times under HSC, more takeovers, and lower maximum steering torque compared to TC, all aligning with our hypotheses (\cref{tab:Hypotheses_metrics_vs_control_modes}).These patterns suggest that HSC keeps drivers more engaged and reduces the need for forceful corrective input. 
Overall, the quantitative results indicate that HSC yields a more favourable MHC profile. 
However, the factors underlying these perceptions, particularly the unexpected positive correlation between reaction time and the perception of sufficient control,remain to be clarified.

Notably, both HSC and TC have relatively high scores for `responsible`, indicating that participants felt morally responsible for the interaction irrespective of the control mode. The qualitative results also mirror this, with participants saying that the control mode defined their exact responsibilities --- with participants expected to supervise in case of TC and correct in case of HSC. The finding that being informed of their responsibility prior to operating the system was sufficient for drivers to internalise it, even when the system had limitations that may have compromised their ability to fulfil it, echoes the concern that drivers of partially automated vehicles were problematically held responsible for accidents even when they were incapable of preventing them~\cite{beckersDriversPartiallyAutomated2022}. These notions of responsibility attribution might change as people get more exposure to capable automation systems that they feel comfortable trusting. To make sense of this ever changing landscape, we distil our findings into a framework for further exploring the perception of MHC in partially automated systems.

\subsection{MHC evaluation framework}
\label{sec:framework-guidelines}

In this work, starting from high-level MHC properties, we followed a hypothesis-driven approach to design an experiment for evaluating objective metrics and subjective perceptions of a human interacting with an automation system in relation to design choices (the control modes).
To aid future research, we summarize the process of designing such experiments (\cref{tab:Steps_for_MHC_exp}).

\begin{table}[th]
    \centering
    \caption{\textbf{Framework for designing meaningful human control evaluation experiments:}
    When studying how humans interact with an automation system, the following steps can be used to design experiments for evaluating their subjective perceptions pertaining to meaningful human control and how they relate to behavioural metrics.}
    \label{tab:Steps_for_MHC_exp}
    \hrule
    \begin{enumerate}[label= Step \arabic*:, leftmargin=*]
    \item Identify a scenario where there is potential conflict between the reasons of the human operator and the automation system which could be caused by mismatches in beliefs, desires or intentions.
    \item Identify the properties of systems under MHC that are relevant to this scenario.
    \item Based on identified properties, identify subjective perceptions to be evaluated in the post-trial surveys and post-experiment questionnaires.
    \item Identify task-relevant objective metrics that pertain to the experiment scenario based on hypothesised relations to subjective perceptions.
    \item Conduct the experiment, collect data, and analyse the results.
    \item (Optional) Revise objective metrics and (questions for) subjective perceptions based on the participant responses to open-ended post-experiment questionnaire.
\end{enumerate}
\hrule
\end{table}

Following the steps in the framework, 
similar experiments could be designed to build a knowledge base of how the subjective perceptions pertaining to MHC are related to different objective metrics. Potential use cases for different target groups are mentioned below.

\noindent\textbf{For automation manufacturers}:
MHC evaluation experiments could be used to direct the design process. Further, results of MHC assessments of final products could be used as evidence in safety cases presented for type approval of AVs by regulatory authorities.

\noindent\textbf{For regulating automated driving systems}:
The framework can be applied to design experiments for evaluating novel designs of driving automation systems from the perspective of the human operators interacting with them.
In case of deficits in MHC, regulatory authorities could request design updates from automation manufacturers or mandate more training for operators.

\noindent\textbf{For assessment of human drivers and remote operators}:
For organisations engaged in the training and licensing of drivers and remote operators, the proposed framework can be used to test the skills of humans in interacting with automation systems. Importantly, the automation systems used in such assessments should be carefully selected to ensure they support MHC, otherwise, the evaluation risks measuring how humans cope with poorly designed automation rather than assessing human drivers' and remote operators' competences.

\subsection{Broader implications for MHC theory}
\label{sec:Dicussion-MHC-framework}
Our methodology adds to the existing literature on evaluating MHC by providing an initial empirical operationalisation
of the established, yet so far mostly abstract, notion of MHC. This is achieved by combining behavioural metrics specifically selected based on hypothesised relationships with MHC-related perceptions of 
the established MHC   properties~\cite{cavalcantesiebertMeaningfulHumanControl2022}) with detailed questionnaires (designed to capture subjective perception of related MHC aspects). Further, although we developed this methodology for evaluating MHC specifically in partially automated driving, it could have broader implications for  
evaluating MHC from the perspective of human operators interacting with a more general class of automated systems.

\subsubsection{Relation to existing MHC evaluation studies}
\label{sec:framework-implications}
This work contributes to the empirical assessment of meaningful human control by translating the four MHC properties~\cite{cavalcantesiebertMeaningfulHumanControl2022} into candidate behavioural metrics selected based on theoretically hypothesised relationships with MHC-related perceptions, linking these indicators to behavioural evidence, and offering a reproducible structure to investigate perceptions of MHC as a function of behavioural metrics and control mode. Beyond the specific comparison of HSC and TC, our findings relate to ongoing debates on complacency, sense of agency, and responsibility in human–automation interaction.

First, we translated the four properties of systems under MHC proposed by \cite{cavalcantesiebertMeaningfulHumanControl2022} into Likert-scale, self-reported quantitative questions. Prior work evaluated MHC mainly through narrative descriptions, with little empirical investigations~\cite{suryana2025meaningful,calvert2020gaps}, without questions explicitly designed to assess MHC. Our approach complements this by employing items that are intentionally constructed to measure the MHC properties.
Another approach quantified MHC for human–robot teams in a computer-based firefighting game based on the tracing condition~\cite{verhagen2024meaningful}. They argue that while subjective measures offer valuable insights into human experiences, objective metrics are essential to verify whether the system genuinely supports MHC. Such objective data helps assess the system's effectiveness in real-time decision-making scenarios and ensures that humans can be held accountable for robot actions, thereby preventing responsibility gaps. 
While their method did not regard the tracking condition, our study is grounded in the four MHC properties~\cite{cavalcantesiebertMeaningfulHumanControl2022} which capture both tracking and tracing conditions. 
Further, in contrast to the simple firefighting game which provides only a simplified representation of complex interactions with discrete actions,
our driving simulator study presents a more ecologically valid scenario with continuous actions and thereby enables a more realistic evaluation of MHC.
Taken together, this positions our contribution as an early empirical operationalisation of MHC in scenarios similar to real-world interactions.

Beyond the operationalisation of MHC introduced above, driver complacency, which can lead to delayed reactions, is a recurring concern in Level 2–3 automation, where drivers might slip into passive monitoring roles \cite{flemischJoiningBluntPointy2019,nordhoff2023mis}.
Our results suggest that HSC may counteract this tendency: participants under HSC showed shorter reaction times and more frequent takeovers than under TC, indicating greater readiness to act. In contrast, TC, with its explicit but infrequent handovers, risks creating phases of over-reliance during which drivers disengage until prompted. Our findings resonate with prior discussions on the two modes of control, where one of the identified pitfalls of TC is that drivers may fall into complacency \cite{abbink2012haptic, flemischJoiningBluntPointy2019,deWinter2023shared}. These patterns suggest that certain interaction designs may undermine meaningful human control, particularly its tracing dimension, as discussed by \cite{calvert2020conceptual}, where the capability to control the system is eroded not due to the absence of human ability, but because the system’s design discourages its exercise. This highlights the importance of designing shared control mechanisms that actively sustain driver engagement, rather than leaving humans in purely supervisory roles.

Beyond engagement, our findings also speak to driver's sense of agency (SoA). Evaluating the SoA of drivers is important as it is directly related to their engagement and how responsible they feel~\cite{mooreWhatSenseAgency2016}. Prior research shows that higher levels of automation often reduce drivers’ SoA, undermining engagement and willingness to accept responsibility~\cite{yunInvestigatingRelationshipAssisted2019,berberianAutomationTechnologySense2012,cornelioSenseAgencyEmerging2022,mooreWhatSenseAgency2016}. In our study, HSC elicited higher ratings of \textit{sufficient control}, \textit{understanding the AV}, and \textit{working together}. These perceptions align conceptually with SoA, suggesting that continuous, transparent, and negotiated interaction helps drivers maintain a subjective feeling of authorship over vehicle behaviour. By contrast, TC can undermine SoA by producing disorientation during abrupt handovers or by diffusing the sense of joint action. 
The qualitative finding that mismatches in intentions can undermine the sense of MHC supports earlier studies that showed how SoA was maintained when automation behaviour matched the intentions of the driver~\cite{wenDecelerationAssistanceMitigated2021}.

The responsibility ascribed to the driver should be commensurate with their agency and, as the level of automation increases, their responsibility should decrease~\cite{flemischDynamicBalanceHumans2012}.
Our results complement prior studies on theoretical and perceived responsibility of humans interacting with automation systems~\cite{douerTheoreticalMeasuredSubjective2021,douerJudgingOneOwn2022} which highlight the importance of considering whether the human can meaningfully contribute to the outcome of the system.
Quantitatively, responsibility ratings did not differ significantly between HSC and TC. Qualitative reports suggest that responsibility was shaped less by control mode itself than by situational demands and drivers’ understanding of their operational role. For example, under HSC, some drivers felt \textit{forced} into responsibility by counter-torque or abrupt interventions, whereas under TC, others felt disengaged or only partially responsible during cooperative execution. These findings highlight that responsibility is not fixed but shaped by how authority is balanced between human and automation: it can be reduced both when the system exerts too much pressure on the driver (dominance) and when it operates too independently, leaving the driver disengaged (independence).
For designers and policymakers, bridging the gap between \emph{assigned} and \emph{experienced} responsibility is essential to prevent unfair attribution of blame when control is shared between humans and automation~\cite{flemischDynamicBalanceHumans2012,methnaniLetMeTake2021,douerJudgingOneOwn2022,beckersDriversPartiallyAutomated2022}.

Finally, although our study focused on automotive shared control, the broader implications extend to other domains where humans and automation jointly act, such as aviation, robotics, and healthcare. Across these domains, complacency~\cite{flemischJoiningBluntPointy2019, nordhoff2023mis},  reduced agency~\cite{cornelioSenseAgencyEmerging2022, berberianAutomationTechnologySense2012}, and blurred responsibility~\cite{beckersDriversPartiallyAutomated2022, 
flemischDynamicBalanceHumans2012} recur as central challenges.
Meaningful human control offers a principled framework for addressing these challenges by guiding the design of systems that sustain engagement, preserve a sense of authorship, and maintain appropriate accountability. In this way, our work connects empirical findings on HSC and TC to broader societal debates on the responsible integration of automation.

\subsubsection{Design guidelines for MHC}
\label{sec:guidelines}

Our evaluation results point to concrete design implications for driving automation systems that aim to comply with the four properties of MHC.

\noindent\textbf{Property 1 (moral operational design domain)}: Drivers require system transparency and familiarisation to interpret automated behaviour. Training sessions, such as those we provided in our experiment, can help users understand system logic across routine and safety-critical scenarios without exposing them to real-world risks. Importantly, participants were explicitly informed that they remained fully responsible for the AV’s actions and were required to stay aware at all times. Such reminders of accountability are essential for aligning drivers’ understanding of their role with the moral ODD of the system.

\noindent\textbf{Property 2 (shared representations)}: 
Mismatches in intentions and difficulty in correcting AV behaviour are detrimental to the realisation of property 2 of MHC. 
Our findings also hint that drivers evaluate automation at the event level rather than aggregated over entire trips. Thus, design should prioritise the quality of individual interactions—for example, ensuring subtle corrections, smooth overrides, trajectory adaptations aligned with driver preferences and improved communication of AV intent through haptic, visual or auditory cues.

\noindent\textbf{Property 3 (sufficient control)}: Because higher trust can delay takeovers, systems should ensure that interventions remain easy and smooth, even after abrupt transitions. Continuous steering access, minimal override friction, and intuitive re-engagement mechanisms are critical design priorities. 

\noindent\textbf{Property 4 (responsibility)}: Responsibility perceptions vary with system design and driving context. Designers should avoid extremes where drivers feel either forced by high conflict or disengaged through passivity. Interfaces and adaptive control strategies that sustain an appropriate sense of accountability, even in low-demand driving, represent a key area for future research.

In summary, our results pertaining to properties 2, 3 and 4, can be distilled into the following guidelines for improving MHC in partially automated driving: 1) facilitate effortless driver interventions, 2) enable transparent communication of automation intent through haptic, visual and auditory cues, and 3) have clear allocation of authority so that drivers are appropriately engaged. 
Together, these implications illustrate how our operationalisation of MHC can not only evaluate but also guide the design of shared-control systems.

\subsection{Limitations and future research}
\label{sec:limitations-future-work}

\subsubsection{Limitations}

The first limitation pertains to the realism of the driving simulator experiment. Driver's perception of risk in the experiment might vary from that during real driving. 
This discrepancy in risk perception might cause participants to behave in a risky manner and to have higher perceptions of safety than in the real case.
Thus, the gap in risk perception between actual and simulated driving limits the generalisability of the results.
Furthermore,
the automation system used in this study is relatively 
artificial in that drivers were only responsible for lateral control and had no input over longitudinal control. This choice was deliberate: by fixing longitudinal control, we isolated the lateral interaction between driver and automation, providing controlled and repeatable conditions for testing meaningful human control. 
However, this simplification limits the generalisability of our findings to real-world driving systems where drivers must manage both lateral and longitudinal control simultaneously, which may alter their perceptions of meaningful human control.

Second, our sample was limited to 24 participants, predominantly students and individuals working in academia, many of whom had limited real-world exposure to automated vehicle technology. This sample size was determined by feasibility constraints and may not fully reflect the perspectives of early adopters or more experienced users of automated systems.
The limited number and homogeneous demographic background of participants limits the generalisability of our findings which may not be representative of the whole population.
However, since our primary contribution is the evaluation methodology rather than definitive conclusions about specific perceptions of MHC, the limited sample size and representativeness does not fundamentally undermine the validity of our approach.

Third, our evaluation focused exclusively on the driver as the human agent, disregarding other relevant human agents such as other road users, traffic regulators, or system designers. While this does not undermine our conclusions, which are scoped to the driver interacting directly with the automation, it does restrict the broader applicability of the findings. MHC only dictates that the system must track the reasons of relevant humans, and that responsibility should be traced back to at least one human. Humans with control and responsibility could be present at different hierarchical levels. For example, in the context of traffic interactions, apart from the driver of a partially automated vehicle, the relevant humans could be the policy makers who set traffic rules, regulatory agencies that certify vehicles, vehicle manufacturers that design automated systems or other traffic participants that interact with the partially automated vehicle.
Future MHC experiments should explore how different humans might assert control across different levels in hierarchies of design and operation.

Fourth,
we used a small number of Likert-scale questions to quantify MHC concepts which are inherently complex and ideally require operationalisation via multi-dimensional constructs.
For example, mutual understanding measured by our questions only partially captures the shared representations property of MHC as perceived by the driver, and leaves out aspects like whether the actual intentions and mental models of the driver matched those of the automation.
Similarly the single question for perception of sufficient control, leaves out aspects like situational awareness and task workload which might affect their ability for exercising control over the automation~\cite{flemischDynamicBalanceHumans2012,cavalcantesiebertMeaningfulHumanControl2022}. 
Finally, the perceived responsibility we measured only concerns the forward link mentioned by Cavalcante Siebert et al.~\cite{cavalcantesiebertMeaningfulHumanControl2022} (awareness of responsibility) but not the backward link (traceability of responsibility).
The choice to focus on a limited number of facets of each property was dictated by practical considerations: some of the aspects excluded from our study would be difficult to measure within our experimental paradigm (e.g., traceability of responsibility), while others could potentially be measured but would have added substantial overhead for our participants (e.g., alignment of representations of surrounding traffic between the human and the automation). 
Further research is needed to develop more comprehensive yet practically feasible evaluations of MHC.

Fifth, most of the behavioural metrics that we hypothesised to be related to MHC-related perceptions were not validated as indicators in this study, as many of the hypothesised relationships were not statistically significant. One possible explanation is the relatively small sample size of 24 participants, which may have limited the statistical power to detect such relationships. Among the hypothesised relationships between behavioural metrics and MHC-related perceptions, only the relationship between conflict in steering torques and perceived AV understanding was confirmed, while the relationship between reaction time and sufficient control was significant but in the opposite direction to our hypothesis. Future studies should therefore further refine and validate these behavioural metrics using larger samples.

Sixth, perceived MHC does not by itself demonstrate that the conditions for MHC have been objectively fulfilled. Drivers may report feeling in control even when their actual ability or authority to influence system outcomes is limited. 
For example,
when evaluating MHC objectively, we would assume that smaller reaction times would offer more MHC, but the subjective scores for \textit{sufficient control} were positively correlated with reaction time.  
Our study represents an initial empirical effort to examine how MHC-related perceptions correspond with observable behavioural evidence, rather than as a direct assessment of whether a system objectively satisfies MHC.

\subsubsection{Future research}

First, in this study, the human outperforms the automation (the AV deliberately fails to steer clear of some road users); an important direction for future research concerns situations in which the automation outperforms the human (e.g., emergency braking or being aware of vehicles in the blind spot).
Such experiments are necessary to identify theoretical bounds for human and automation performance with regard to different levels of driving complexity~\cite{dewinterSharedControlTraded2023}, and for designing appropriate transfers of authority.

Second, the relatively short duration of the experiment cannot capture how driving behaviour might adapt over days of exposure to partial automation. These adaptations in human behaviour along with updates in automation behaviour can lead to unforeseen emergent behaviours that could be detrimental to meaningful human control~\cite{melmanMitigatingUndesirableEmergent2020,calvertDesigningAutomatedVehicle2023}. Thus, to manage emergence, longer studies that account for adaptation are needed to properly validate automated systems. 

Third, future studies should include a broader and more diverse participant pool, such as incorporating drivers with varied ages, professional backgrounds, driving experience, and geographical locations, to provide additional insights into how perceptions of MHC vary across different user groups.

Fourth, more continuous information about driver perception could be obtained by measuring physiological signals such as heart rate, blood pressure, muscle activation, or brain activity, for example, relating the sense of agency to electroencephalography (EEG) signals~\cite{yunInvestigatingRelationshipAssisted2019}.

Fifth, extending the evaluation framework to include other human agents with varying degrees of authority and influence, such as other road users, traffic regulators, or system designers, would provide a more complete picture of MHC in partially automated driving.

\section{Conclusion}

Starting from the four properties of systems under meaningful human control (MHC), we developed and demonstrated an MHC evaluation methodology, instantiated here through a driving simulator experiment, to study the perception of responsibility and control in drivers of partially automated vehicles operating under two control modes --- haptic shared control (HSC) and traded control (TC). Although we used simplified implementations of the control modes (without important features such as adaptive references, auditory warnings, or take-over requests) they elicited differing perceptions and interaction experiences, serving as a testing ground for our MHC evaluation methodology.
We integrated behavioural metrics from telemetry data, subjective perception scores from post-trial surveys and qualitative feedback from open-ended post-trial questionnaires. 
Linear mixed-effect models confirmed our hypothesis that conflict in steering torques is negatively correlated with the perception of the AV understanding the driver. 
However, contrary to our hypothesis, the analysis revealed a surprising positive correlation between reaction time and the perception of sufficient control which contradicted our hypothesis. 
Such contradictions highlight the need for conducting experiments for evaluating design choices regarding the interaction between driver and automated driving systems.

From the qualitative feedback, we identified three categories of factors --- attitudinal factors, interaction factors and trajectory factors --- that positively and negatively influence the perception of MHC. 
Mismatches in intentions, lack of safety and resistance to driver inputs were frequently associated with reduction in perceived MHC. The perception of being responsible on the other hand was more dependent on the expectations associated with being accountable for the novel and untrustworthy automation, especially in safety-critical situations and less dependent on the control mode.
Finally the exploratory analysis suggested that HSC supports meaningful human control better than TC; specifically the results for the perceptions of `sufficient control', `understanding the AV', and `working together'.

Based on our findings, guidelines for designing the interaction of drivers with the automation are summarised as: (1) ensure that driver interventions remain effortless and smooth, (2) improve mutual communication of intentions between the driver and the automation, and (3) ensure clear allocation of authority to ensure that drivers are appropriately engaged.
Ultimately, we presented a framework for designing experiments starting from the properties of systems under MHC,
to guide future research on the perception of MHC for humans interacting with automated systems.

\section*{CRediT author statement}

\textbf{Conceptualization}: AG, LES, LF, DA, SCC, LCS, AZ;
\textbf{Methodology}: AG, LES, LF, SCC, LCS, AZ;
\textbf{Software}: AG, LES, LF;
\textbf{Validation}: AG, LES;
\textbf{Formal analysis}: AG, LES. LF;
\textbf{Investigation}: AG, LES, LF;
\textbf{Resources}: DA, AZ;
\textbf{Data curation}: AG, LES;
\textbf{Writing - original draft preparation}: AG, LES;
\textbf{Writing - review and editing}: AG, LES, BvA, DA, SCC, LCS, AZ;
\textbf{Visualization}: AG, LES;
\textbf{Supervision}: BvA, DA, SCC, LCS, AZ;

\section*{Acknowledgements}

This research was funded by the TU Delft AI Initiative and Indonesia Endowment Fund for Education Agency (LPDP). The authors would like to thank members of the Centre for Meaningful Human Control and  AiTech for all their valuable suggestions, feedback and discussions during our meetings.

\section*{Declaration of generative AI and AI-assisted technologies in the manuscript preparation process.}

During the preparation of this work the authors used ChatGPT from OpenAI in order to enhance clarity and conciseness of the manuscript. After using this tool, the authors reviewed and edited the content as needed and take full responsibility for the content of the published article.

\small
\bibliography{Emergence_And_Responsibility_AG,Emergence_And_Responsibility}

\begin{thebibliography}{10}

\bibitem{flemischDynamicBalanceHumans2012}
F.~Flemisch, M.~Heesen, T.~Hesse, J.~Kelsch, A.~Schieben, and J.~Beller, ``Towards a dynamic balance between humans and automation: Authority, ability, responsibility and control in shared and cooperative control situations,'' {\em Cognition, Technology \& Work}, vol.~14, pp.~3--18, Mar. 2012.

\bibitem{beckersDriversPartiallyAutomated2022}
N.~Beckers, L.~C. Siebert, M.~Bruijnes, C.~Jonker, and D.~Abbink, ``Drivers of partially automated vehicles are blamed for crashes that they cannot reasonably avoid,'' {\em Scientific Reports}, vol.~12, p.~16193, Sept. 2022.

\bibitem{fagnant2015}
D.~J. Fagnant and K.~Kockelman, ``Preparing a nation for autonomous vehicles: opportunities, barriers and policy recommendations,'' {\em Transportation Research Part A: Policy and Practice}, vol.~77, pp.~167--181, 2015.

\bibitem{kusano2025waymo}
K.~D. Kusano, J.~M. Scanlon, Y.~H. Chen, T.~L. McMurry, T.~Gode, and T.~Victor, ``Comparison of waymo rider-only crash rates by crash type to human benchmarks at 56.7 million miles,'' {\em Traffic Injury Prevention}, pp.~1--13, 2025.

\bibitem{kim2026comparative}
Y.~Kim, S.~Kim, D.~Kim, and K.~Lee, ``Comparative review of commercialized advanced driver assistance system (adas) technologies,'' {\em Electronics}, vol.~15, no.~10, p.~2015, 2026.

\bibitem{sae2021taxonomy}
{SAE International}, ``Taxonomy and definitions for terms related to on-road motor vehicle automated driving systems,'' Standard J3016\_202104, 2021.

\bibitem{endsley2017autonomous}
M.~R. Endsley, ``Autonomous driving systems: A preliminary naturalistic study of the tesla model s,'' {\em Journal of Cognitive Engineering and Decision Making}, vol.~11, no.~3, pp.~225--238, 2017.

\bibitem{chu2023automation}
Y.~Chu and P.~Liu, ``Automation complacency on the road,'' {\em Ergonomics}, vol.~66, no.~11, pp.~1730--1749, 2023.

\bibitem{merat2014human}
N.~Merat, H.~A.~Jamson, F.~Lai, and O.~Carsten, ``Human factors of highly automated driving: results from the easy and citymobil projects,'' in {\em Road vehicle automation}, pp.~113--125, Springer, 2014.

\bibitem{moore2016sense}
J.~W. Moore, ``What is the sense of agency and why does it matter?,'' {\em Frontiers in psychology}, vol.~7, p.~1272, 2016.

\bibitem{cornelio2022sense}
P.~Cornelio, P.~Haggard, K.~Hornbaek, O.~Georgiou, J.~Bergstr{\"o}m, S.~Subramanian, and M.~Obrist, ``The sense of agency in emerging technologies for human--computer integration: A review,'' {\em Frontiers in Neuroscience}, vol.~16, p.~949138, 2022.

\bibitem{berberian2012automation}
B.~Berberian, J.-C. Sarrazin, P.~Le~Blaye, and P.~Haggard, ``Automation technology and sense of control: a window on human agency,'' {\em PloS one}, vol.~7, no.~3, p.~e34075, 2012.

\bibitem{moretto2011experience}
G.~Moretto, E.~Walsh, and P.~Haggard, ``Experience of agency and sense of responsibility,'' {\em Consciousness and cognition}, vol.~20, no.~4, pp.~1847--1854, 2011.

\bibitem{matthias2004responsibility}
A.~Matthias, ``The responsibility gap: Ascribing responsibility for the actions of learning automata,'' {\em Ethics and information technology}, vol.~6, pp.~175--183, 2004.

\bibitem{santoni2021four}
F.~Santoni~de Sio and G.~Mecacci, ``Four responsibility gaps with artificial intelligence: Why they matter and how to address them,'' {\em Philosophy \& Technology}, vol.~34, no.~4, pp.~1057--1084, 2021.

\bibitem{calvert2020gaps}
S.~C. Calvert, B.~van Arem, D.~D. Heikoop, M.~Hagenzieker, G.~Mecacci, and F.~S. de~Sio, ``Gaps in the control of automated vehicles on roads,'' {\em IEEE intelligent transportation systems magazine}, vol.~13, no.~4, pp.~146--153, 2020.

\bibitem{santoni2018meaningful}
F.~Santoni~de Sio and J.~Van~den Hoven, ``Meaningful human control over autonomous systems: A philosophical account,'' {\em Frontiers in Robotics and AI}, vol.~5, p.~15, 2018.

\bibitem{cavalcantesiebertMeaningfulHumanControl2022}
L.~Cavalcante~Siebert, M.~L. Lupetti, E.~Aizenberg, N.~Beckers, A.~Zgonnikov, H.~Veluwenkamp, D.~Abbink, E.~Giaccardi, G.-J. Houben, C.~M. Jonker, J.~{van den Hoven}, D.~Forster, and R.~L. Lagendijk, ``Meaningful human control: Actionable properties for {{AI}} system development,'' {\em AI and Ethics}, May 2022.

\bibitem{calvert2025principles}
S.~C. Calvert, ``Principles and framework for the operationalisation of meaningful human control over autonomous systems,'' {\em Science and Engineering Ethics}, vol.~31, no.~5, p.~27, 2025.

\bibitem{calvertDesigningAutomatedVehicle2023}
S.~C. Calvert, S.~O. Johnsen, and A.~George, ``Designing automated vehicle and traffic systems towards meaningful human control,'' in {\em Research Handbook on Meaningful Human Control of Artificial Intelligence Systems}, United Kingdom: Edward Elgar Publishing, 2023 (Accepted).

\bibitem{mecacci2020meaningful}
G.~Mecacci and F.~Santoni~de Sio, ``Meaningful human control as reason-responsiveness: the case of dual-mode vehicles,'' {\em Ethics and Information Technology}, vol.~22, no.~2, pp.~103--115, 2020.

\bibitem{vanderwaaAllocationMoralDecisionMaking2020}
J.~Van Der~Waa, J.~Van~Diggelen, L.~Cavalcante~Siebert, M.~Neerincx, and C.~Jonker, ``Allocation of {{Moral Decision-Making}} in {{Human-Agent Teams}}: {{A Pattern Approach}},'' in {\em Engineering {{Psychology}} and {{Cognitive Ergonomics}}. {{Cognition}} and {{Design}}} (D.~Harris and W.-C. Li, eds.), vol.~12187, pp.~203--220, Cham: Springer International Publishing, 2020.

\bibitem{suryana2025meaningful}
L.~E. Suryana, S.~Nordhoff, S.~Calvert, A.~Zgonnikov, and B.~van Arem, ``Meaningful human control of partially automated driving systems: Insights from interviews with tesla users,'' {\em Transportation Research Part F: Traffic Psychology and Behaviour}, vol.~113, pp.~213--236, 2025.

\bibitem{verhagen2024meaningful}
R.~S. Verhagen, M.~A. Neerincx, and M.~L. Tielman, ``Meaningful human control and variable autonomy in human-robot teams for firefighting,'' {\em Frontiers in Robotics and AI}, vol.~11, p.~1323980, 2024.

\bibitem{abbink2018topology}
D.~A. Abbink, T.~Carlson, M.~Mulder, J.~C. De~Winter, F.~Aminravan, T.~L. Gibo, and E.~R. Boer, ``A topology of shared control systems—finding common ground in diversity,'' {\em IEEE Transactions on Human-Machine Systems}, vol.~48, no.~5, pp.~509--525, 2018.

\bibitem{wang2017effect}
Z.~Wang, R.~Zheng, T.~Kaizuka, K.~Shimono, and K.~Nakano, ``The effect of a haptic guidance steering system on fatigue-related driver behavior,'' {\em IEEE Transactions on Human-Machine Systems}, vol.~47, no.~5, pp.~741--748, 2017.

\bibitem{li2018shared}
M.~Li, H.~Cao, X.~Song, Y.~Huang, J.~Wang, and Z.~Huang, ``Shared control driver assistance system based on driving intention and situation assessment,'' {\em IEEE Transactions on Industrial Informatics}, vol.~14, no.~11, pp.~4982--4994, 2018.

\bibitem{deWinter2023shared}
J.~C. de~Winter, S.~M. Petermeijer, and D.~A. Abbink, ``Shared control versus traded control in driving: a debate around automation pitfalls,'' {\em Ergonomics}, vol.~66, no.~10, pp.~1494--1520, 2023.

\bibitem{beckers2023joan}
N.~Beckers, O.~Siebinga, J.~Giltay, and A.~van~der Kraan, ``Joan: A framework for human-automated vehicle interaction experiments in a virtual reality driving simulator,'' {\em Journal of Open Source Software}, vol.~8, no.~82, p.~4250, 2023.

\bibitem{dosovitskiy2017carla}
A.~Dosovitskiy, G.~Ros, F.~Codevilla, A.~Lopez, and V.~Koltun, ``Carla: An open urban driving simulator,'' in {\em Conference on robot learning}, pp.~1--16, PMLR, 2017.

\bibitem{van2017four}
M.~R. van Paassen, R.~P. Boink, D.~A. Abbink, M.~Mulder, and M.~Mulder, ``Four design choices for haptic shared control,'' in {\em Advances in Aviation Psychology, Volume 2}, pp.~237--254, Routledge, 2017.

\bibitem{sheridan1978human}
T.~B. Sheridan, W.~L. Verplank, and T.~Brooks, ``Human/computer control of undersea teleoperators,'' in {\em NASA. Ames Res. Center The 14th Ann. Conf. on Manual Control}, 1978.

\bibitem{remyLearningResponsibilityAllocations2024}
I.~Remy, D.~{Fridovich-Keil}, and K.~Leung, ``Learning responsibility allocations for multi-agent interactions: {{A}} differentiable optimization approach with control barrier functions,'' Oct. 2024.

\bibitem{vincentMoralResponsibility2011}
N.~A. Vincent, I.~{van de Poel}, and J.~{van den Hoven}, eds., {\em Moral {{Responsibility}}}, vol.~27 of {\em Library of {{Ethics}} and {{Applied Philosophy}}}.
\newblock Dordrecht: Springer Netherlands, 2011.

\bibitem{suryana2024meaningful}
L.~E. Suryana, S.~Nordhoff, S.~C. Calvert, A.~Zgonnikov, and B.~Van~Arem, ``A meaningful human control perspective on user perception of partially automated driving systems: a case study of tesla users,'' in {\em 2024 IEEE intelligent vehicles symposium (IV)}, pp.~409--416, IEEE, 2024.

\bibitem{pengConceptualisingUserComfort2024}
C.~Peng, S.~Horn, R.~Madigan, C.~Marberger, J.~D. Lee, J.~Krems, M.~Beggiato, R.~Romano, C.~Wei, E.~Wooldridge, R.~Happee, M.~Hagenzieker, and N.~Merat, ``Conceptualising user comfort in automated driving: {{Findings}} from an expert group workshop,'' {\em Transportation Research Interdisciplinary Perspectives}, vol.~24, p.~101070, Mar. 2024.

\bibitem{korberTheoreticalConsiderationsDevelopment2019}
M.~K{\"o}rber, ``Theoretical {{Considerations}} and {{Development}} of a {{Questionnaire}} to {{Measure Trust}} in {{Automation}},'' in {\em Proceedings of the 20th {{Congress}} of the {{International Ergonomics Association}} ({{IEA}} 2018)} (S.~Bagnara, R.~Tartaglia, S.~Albolino, T.~Alexander, and Y.~Fujita, eds.), vol.~823, pp.~13--30, Cham: Springer International Publishing, 2019.

\bibitem{nordhoffPerceivedSafetyTrust2021}
S.~Nordhoff, J.~Stapel, X.~He, A.~Gentner, and R.~Happee, ``Perceived safety and trust in {{SAE Level}} 2 partially automated cars: {{Results}} from an online questionnaire,'' {\em PLOS ONE}, vol.~16, p.~e0260953, Dec. 2021.

\bibitem{boink2014understanding}
R.~Boink, M.~M. Van~Paassen, M.~Mulder, and D.~A. Abbink, ``Understanding and reducing conflicts between driver and haptic shared control,'' in {\em 2014 IEEE International Conference on Systems, Man, and Cybernetics (SMC)}, pp.~1510--1515, IEEE, 2014.

\bibitem{ercan2018predictive}
Z.~Ercan, A.~Carvalho, H.~E. Tseng, M.~G{\"o}ka{\c{s}}an, and F.~Borrelli, ``A predictive control framework for torque-based steering assistance to improve safety in highway driving,'' {\em Vehicle system dynamics}, vol.~56, no.~5, pp.~810--831, 2018.

\bibitem{mars2014analysis}
F.~Mars, M.~Deroo, and J.-M. Hoc, ``Analysis of human-machine cooperation when driving with different degrees of haptic shared control,'' {\em IEEE transactions on haptics}, vol.~7, no.~3, pp.~324--333, 2014.

\bibitem{markkula2006steering}
G.~Markkula and J.~Engstr{\"o}m, ``A steering wheel reversal rate metric for assessing effects of visual and cognitive secondary task load,'' in {\em Proceedings of the 13th ITS World Congress}, Leeds, 2006.

\bibitem{jiao2023ttc}
Y.~Jiao, ``{A fast calculation of two-dimensional Time-to-Collision},'' Mar. 2023.

\bibitem{cohen1960coefficient}
J.~Cohen, ``A coefficient of agreement for nominal scales,'' {\em Educational and psychological measurement}, vol.~20, no.~1, pp.~37--46, 1960.

\bibitem{gwet2008computing}
K.~L. Gwet, ``Computing inter-rater reliability and its variance in the presence of high agreement,'' {\em British Journal of Mathematical and Statistical Psychology}, vol.~61, no.~1, pp.~29--48, 2008.

\bibitem{braun2006using}
V.~Braun and V.~Clarke, ``Using thematic analysis in psychology,'' {\em Qualitative research in psychology}, vol.~3, no.~2, pp.~77--101, 2006.

\bibitem{hill2005consensual}
C.~E. Hill, B.~J. Thompson, and E.~N. Williams, ``Consensual qualitative research: An update,'' {\em Journal of Counseling Psychology}, vol.~52, no.~2, pp.~196--205, 2005.

\bibitem{campbell2013coding}
J.~L. Campbell, C.~Quincy, J.~Osserman, and O.~K. Pedersen, ``Coding in-depth semistructured interviews: Problems of unitization and intercoder reliability and agreement,'' {\em Sociological methods \& research}, vol.~42, no.~3, pp.~294--320, 2013.

\bibitem{feinstein1990high}
A.~R. Feinstein and D.~V. Cicchetti, ``High agreement but low kappa: I. the problems of two paradoxes,'' {\em Journal of clinical epidemiology}, vol.~43, no.~6, pp.~543--549, 1990.

\bibitem{payre2016fully}
W.~Payre, J.~Cestac, and P.~Delhomme, ``Fully automated driving: Impact of trust and practice on manual control recovery,'' {\em Human factors}, vol.~58, no.~2, pp.~229--241, 2016.

\bibitem{dixit2016autonomous}
V.~V. Dixit, S.~Chand, and D.~J. Nair, ``Autonomous vehicles: disengagements, accidents and reaction times,'' {\em PLoS one}, vol.~11, no.~12, p.~e0168054, 2016.

\bibitem{wenSenseAgencyPerception2022}
W.~Wen and H.~Imamizu, ``The sense of agency in perception, behaviour and human--machine interactions,'' {\em Nature Reviews Psychology}, vol.~1, pp.~211--222, Feb. 2022.

\bibitem{berberianAutomationTechnologySense2012}
B.~Berberian, J.-C. Sarrazin, P.~Le~Blaye, and P.~Haggard, ``Automation {{Technology}} and {{Sense}} of {{Control}}: {{A Window}} on {{Human Agency}},'' {\em PLoS ONE}, vol.~7, p.~e34075, Mar. 2012.

\bibitem{baysAttenuationSelfGeneratedTactile2006}
P.~M. Bays, J.~R. Flanagan, and D.~M. Wolpert, ``Attenuation of {{Self-Generated Tactile Sensations Is Predictive}}, not {{Postdictive}},'' {\em PLoS Biology}, vol.~4, p.~e28, Jan. 2006.

\bibitem{wenkeHowVoluntaryActions2009}
D.~Wenke and P.~Haggard, ``How voluntary actions modulate time perception,'' {\em Experimental Brain Research}, vol.~196, pp.~311--318, July 2009.

\bibitem{georgeFeasibleActionSpace2025}
A.~George, L.~C. Siebert, D.~A. Abbink, and A.~Zgonnikov, ``Feasible {{Action Space Reduction}} for {{Quantifying Causal Responsibility}} in {{Continuous Spatial Interactions}},'' May 2025.

\bibitem{shalev-shwartzVisionZeroProvable2019}
S.~{Shalev-Shwartz}, S.~Shammah, and A.~Shashua, ``Vision {{Zero}}: On a {{Provable Method}} for {{Eliminating Roadway Accidents}} without {{Compromising Traffic Throughput}},'' Jan. 2019.

\bibitem{chenEvaluatingSafetyEfficiency2024}
K.~Chen, Z.~Li, P.~Liu, V.~L. Knoop, Y.~Han, and Y.~Jiao, ``Evaluating the safety and efficiency impacts of forced lane change with negative gaps based on empirical vehicle trajectories,'' {\em Accident Analysis \& Prevention}, vol.~203, p.~107622, Aug. 2024.

\bibitem{papadimitriouCommonEthicalSafe2022}
E.~Papadimitriou, H.~Farah, G.~{van de Kaa}, F.~{Santoni de Sio}, M.~Hagenzieker, and P.~{van Gelder}, ``Towards common ethical and safe `behaviour' standards for automated vehicles,'' {\em Accident Analysis \& Prevention}, vol.~174, p.~106724, Sept. 2022.

\bibitem{nordhoff2023mis}
S.~Nordhoff, J.~D. Lee, S.~C. Calvert, S.~Berge, M.~Hagenzieker, and R.~Happee, ``(mis-) use of standard autopilot and full self-driving (fsd) beta: Results from interviews with users of tesla's fsd beta,'' {\em Frontiers in psychology}, vol.~14, p.~1101520, 2023.

\bibitem{molnar2018understanding}
L.~J. Molnar, L.~H. Ryan, A.~K. Pradhan, D.~W. Eby, R.~M.~S. Louis, and J.~S. Zakrajsek, ``Understanding trust and acceptance of automated vehicles: An exploratory simulator study of transfer of control between automated and manual driving,'' {\em Transportation research part F: traffic psychology and behaviour}, vol.~58, pp.~319--328, 2018.

\bibitem{flemischJoiningBluntPointy2019}
F.~Flemisch, D.~A. Abbink, M.~Itoh, M.-P. {Pacaux-Lemoine}, and G.~We{\ss}el, ``Joining the blunt and the pointy end of the spear: Towards a common framework of joint action, human--machine cooperation, cooperative guidance and control, shared, traded and supervisory control,'' {\em Cognition, Technology \& Work}, vol.~21, pp.~555--568, Nov. 2019.

\bibitem{abbink2012haptic}
D.~A. Abbink, M.~Mulder, and E.~R. Boer, ``Haptic shared control: smoothly shifting control authority?,'' {\em Cognition, Technology \& Work}, vol.~14, pp.~19--28, 2012.

\bibitem{calvert2020conceptual}
S.~C. Calvert and G.~Mecacci, ``A conceptual control system description of cooperative and automated driving in mixed urban traffic with meaningful human control for design and evaluation,'' {\em IEEE Open Journal of Intelligent Transportation Systems}, vol.~1, pp.~147--158, 2020.

\bibitem{mooreWhatSenseAgency2016}
J.~W. Moore, ``What {{Is}} the {{Sense}} of {{Agency}} and {{Why Does}} it {{Matter}}?,'' {\em Frontiers in Psychology}, vol.~7, Aug. 2016.

\bibitem{yunInvestigatingRelationshipAssisted2019}
S.~Yun, W.~Wen, Q.~An, S.~Hamasaki, H.~Yamakawa, Y.~Tamura, A.~Yamashita, and H.~Asama, ``Investigating the relationship between assisted driver's {{SoA}} and {{EEG}},'' in {\em Converging Clinical and Engineering Research on Neurorehabilitation {{III}}} (L.~Masia, S.~Micera, M.~Akay, and J.~L. Pons, eds.), (Cham), pp.~1039--1043, Springer International Publishing, 2019.

\bibitem{cornelioSenseAgencyEmerging2022}
P.~Cornelio, P.~Haggard, K.~Hornbaek, O.~Georgiou, J.~Bergstr{\"o}m, S.~Subramanian, and M.~Obrist, ``The sense of agency in emerging technologies for human--computer integration: {{A}} review,'' {\em Frontiers in Neuroscience}, vol.~16, Sept. 2022.

\bibitem{wenDecelerationAssistanceMitigated2021}
W.~Wen, S.~Yun, A.~Yamashita, B.~D. Northcutt, and H.~Asama, ``Deceleration {{Assistance Mitigated}} the {{Trade}}-off {{Between Sense}} of {{Agency}} and {{Driving Performance}},'' {\em Frontiers in Psychology}, vol.~12, June 2021.

\bibitem{douerTheoreticalMeasuredSubjective2021}
N.~Douer and J.~Meyer, ``Theoretical, {{Measured}}, and {{Subjective Responsibility}} in {{Aided Decision Making}},'' {\em ACM Transactions on Interactive Intelligent Systems}, vol.~11, pp.~1--37, Mar. 2021.

\bibitem{douerJudgingOneOwn2022}
N.~Douer and J.~Meyer, ``Judging {{One}}'s {{Own}} or {{Another Person}}'s {{Responsibility}} in {{Interactions With Automation}},'' {\em Human Factors: The Journal of the Human Factors and Ergonomics Society}, vol.~64, pp.~359--371, Mar. 2022.

\bibitem{methnaniLetMeTake2021}
L.~Methnani, A.~Aler~Tubella, V.~Dignum, and A.~Theodorou, ``Let {{Me Take Over}}: {{Variable Autonomy}} for {{Meaningful Human Control}},'' {\em Frontiers in Artificial Intelligence}, vol.~4, p.~737072, Sept. 2021.

\bibitem{dewinterSharedControlTraded2023}
J.~C.~F. De~Winter, S.~M. Petermeijer, and D.~A. Abbink, ``Shared control versus traded control in driving: A debate around automation pitfalls,'' {\em Ergonomics}, vol.~66, pp.~1494--1520, Oct. 2023.

\bibitem{melmanMitigatingUndesirableEmergent2020}
T.~Melman, N.~Beckers, and D.~Abbink, ``Mitigating undesirable emergent behavior arising between driver and semi-automated vehicle,'' {\em arXiv preprint arXiv:2006.16572}, 2020.

\end{thebibliography}
\normalsize

\end{document}